\documentclass[onecolumn,amsmath,amssymb,nofootinbib,12pt]{article}
\usepackage{jheppub}

\usepackage{graphicx}
\usepackage{dcolumn}
\usepackage{bm,psfrag}

\usepackage{subfigure}
\usepackage{float}
\usepackage{tensor}

\newcommand{\ben}{\begin{equation}}
\newcommand{\een}{\end{equation}}
\newcommand{\be}{\begin{equation}}
\newcommand{\ee}{\end{equation}}
\newcommand{\bea}{\begin{eqnarray}}
\newcommand{\eea}{\end{eqnarray}}
\newcommand{\ba}{\begin{eqnarray}}
\newcommand{\ea}{\end{eqnarray}}

\newcommand{\beq}{\begin{equation}}
\newcommand{\eeq}{\end{equation}}
\newcommand{\beqa}{\begin{eqnarray}}
\newcommand{\eeqa}{\end{eqnarray}}
\newcommand{\beqar}{\begin{eqnarray*}}
\newcommand{\eeqar}{\end{eqnarray*}}

\newcommand{\cO}{{\cal O}}

\def\t6 {T_\mt{D6}}

\newcommand{\mt}[1]{\textrm{\tiny #1}}

\newcommand{\ha}{H_{\!\ssc A}}

\newcommand{\vx}{{\vec{x}}}

\def\cC{{\cal C}}

\def\cale         {{\cal E}}

\def\calk         {{\cal K}}

\def\calp         {{\cal P}}

\def\tx{{\tilde{x}}}

\def\ee           {{\rm e}}

\def\Re           {{\rm Re\hskip0.1em}}

\def\bz{\bar{z}}
\def\brho{\bar{\rho}}
\def\trho{\tilde{\rho}}

\def\sqr#1#2{{\vcenter{\vbox{\hrule height.#2pt
 \hbox{\vrule width.#2pt height#1pt \kern#1pt
 \vrule width.#2pt}\hrule height.#2pt}}}}

\def\ee{\cale}
\def\hx{\hat{x}}

\def\hA{\hat{A}}

\def\hp{\hat{p}}

\def\hb{\hat{b}}
\def\hu{\hat{u}}

\def\hN{\hat{N}}
\def\hpsi{\hat{\psi}}
\def\hc{\hat{c}}

\def\aa1{\phi}
\def\cc1{\psi}

\def\hchi{\hat{\chi}}
\def\hxi{\hat{\xi}}
\def\hlambda{\hat{\lambda}}

\def\ha{\hat{a}}

\def\vp{\vec{p}}
\def\vx{\vec{x}}
\def\vy{\vec{y}}
\def\vq{\vec{q}}

\def\vw{\vec{w}}

\begin{document}


\title{Entanglement Entropy and Phase Space Density: Lowest Landau Levels and 1/2 BPS states}

\author{Sumit R. Das $^1$}
\author{Shaun Hampton $^2$}
\author{Sinong Liu $^3$}
\affiliation{$1$ Department of Physics and Astronomy, University of Kentucky,\\ 
\vphantom{k}\ \ Lexington, KY 40506, USA}
\affiliation{$2$ 
Institut de Physique Theorique,
Universite Paris Saclay, CEA, CNRS,\\
Orme des Merisiers, Gif sur Yvette, 91191 CEDEX, France}
{\affiliation{$3$ Faculty of Physics, University of Warsaw, ul. Pasteura 5, 02-093 Warsaw, Poland.}


\date{\today}

\abstract{We consider the entanglement entropy of an arbitrary subregion in a system of $N$ non-relativistic fermions in $2+1$ dimensions in Lowest Landau Level (LLL) states. Using the connection of these states to those of an auxiliary $1+1$ dimensional fermionic system, we derive an expression for the leading large-$N$ contribution in terms of the expectation value of the phase space density operator in $1+1$ dimensions. For appropriate subregions the latter can replaced by its semiclassical Thomas-Fermi value, yielding expressions in terms of explicit integrals which can be evaluated analytically. We show that the leading term in the entanglement entropy is a perimeter law with a shape independent coefficient. Furthermore, we obtain analytic expressions for additional contributions from sharp corners on the entangling curve. Both the perimeter and the corner pieces are in good agreement with existing calculations for special subregions. Our results are relevant to the integer quantum Hall effect problem, and to the half-BPS sector of $\mathcal N=4$ Yang Mills theory on $S^3$. In this latter context, the entanglement we consider is an entanglement in target space. We comment on possible implications to gauge-gravity duality.}

\maketitle


\newpage

\section{Introduction}

It is well known that a large number of free 
non-relativistic fermions in an external potential can be described in the semiclassical limit in
terms of a fluid in phase space. In this limit, the
phase space density operator $\hu(\vp,\vx,t)$ expressed in terms of the fermion field $\hpsi(\vx)$,
\ben
\hu(\vp,\vx,t) = \int d^d\vy~\hpsi^\dagger \bigg(\vx-\frac{\vy}{2},t\bigg)\hpsi \bigg(\vx + \frac{\vy}{2},t\bigg)~e^{i\vp \cdot \vy/\hbar} 
\label{2-8}
\een
becomes {\em classical}, i.e. its expectation value is zero or
one. This is the Thomas-Fermi approximation \cite{thomasfermi}. For example, the ground state for some time independent hamiltonian $H(\vp,\vx)$ is typically described by a single connected region in phase space bounded by the curve $H(\vp,\vx)=E_0$. This simplification becomes particularly useful in one spatial dimension.
In this context, the phase space picture has been applied e.g. in the $c=1$ Matrix Model \cite{bipz}, and has been used \cite{polchinski} to understand the classical limit of collective field theory \cite{jevsak} \footnote{There are also suggestions that $\hu(\vp,\vx,t)$ might be useful beyond this classical limit \cite{ddmw,dmw}. In this paper, we will deal entirely with the Thomas-Fermi approximation.}.

In \cite{klich} it has been shown, using the results of \cite{peschel}, that in free fermion field theory the entanglement entropy of a subregion can be expressed as a sum over cumulants of the particle number distribution. For a large number of fermions the leading term in many systems is the lowest cumulant - dispersion of the number of fermions in the subregion. This appears to be a good approximation for our system (see e.g. \cite{klichqhe}), even though the higher order cumulants are not parametrically suppressed (see also \cite{french}) \footnote{ See below for more comments on this issue.}. For slater determinant states, the cumulants can be expressed in terms of integrals of two point correlators of fermions \cite{cala3,calsatya}, and therefore in terms of fourier transforms of the expectation value of the phase space density \cite{dhl1}. In the Thomas-Fermi approximation, the latter can be replaced by its classical value. This approximation leads to a fermion two point function which is reliable in the ``bulk regime'' and at short distances \cite{satya3,satya4,satya2} \footnote{We thank A. Jevicki for raising a question which clarified the regime of validity of this approximation, and S. Majumdar for bringing \cite{satya3,satya4}, which contain discussions of the validity, to our attention.}. Therefore, for appropriate intervals, this leads to a simple expression for the entanglement entropy. This approximation is also relevant for excited states which are described by filled regions of the single particle phase space \cite{satya4} which can be in general disconnected.

In this paper, we will use this observation to calculate the
entanglement entropy of a system of {\em free} $2+1$ dimensional
fermions $\psi(x_1,x_2,t)$ in Lowest Landau Level (LLL) states. These
states appear in many problems, most famously in the problem of
charged particles moving in a constant external magnetic field. In
this context, the problem is that of integer quantum Hall
effect with $\nu =1$.
Another example involves the holomorphic sector of free $2+1$
dimensional fermions in the presence of an isotropic harmonic oscillator
potential. One interesting context where the latter system appears is in the discussion of the 1/2-BPS sector of $\mathcal N=4$ Yang-Mills theory on $S^3$. 
As shown in \cite{antal}, and discussed extensively in the literature
(see e.g. \cite{berenstein,tsuchiya,ghodsi}), these states can be
mapped to singlet holomorphic states of a complex matrix model which
results from the reduction on $S^3$. For a gauge group $SU(N)$, these
can be further described by the holomorphic sector of $N$ fermions in $2+1$ dimensions in an external isotropic harmonic oscillator potential. 
In addition, there are $\frac{1}{2}N(N-1)$ bosons which are always in their ground state. The holomorphicity
condition effectively relates the system to a one dimensional harmonic oscillator - which is the auxiliary one dimensional system. The fermion wavefunctions are in one-to-one correspondence with Schur polynomials of $SU(N)$ \cite{antal2}.
There are extensions of this map for states with lower supersymmetry
\cite{donos, mandal2}.

{In this latter setup, our computation is that of a {\em target space entanglement entropy} of the type discussed in  \cite{dmt1,dmt2,mazenc,lawrence,sotaro,hartnoll2} in the complex matrix model in this special sector}. A key motivation for our work is to use this connection to understand target space entanglement in the context of AdS/CFT
correspondence. There is an intimate relationship between entanglement
in the base space of quantum field theories and the emergence of
smooth gravitational duals \cite{enta}. It is natural to believe that
there is a similar connection involving target space entanglement,
since parts of the gravitational background arise from the target
space of the field theory (e.g. the $S^5$ in $AdS_5 \times S^5$). For
holographic correspondences like the $c=1$ Matrix Model/two
dimensional string theory or the BFSS Matrix Theory/M-theory dualities, the dual theory is
quantum mechanics of matrices, and the entanglement structure of the
state is entirely target space entanglement. In fact, the calculation
of target space entanglement in the $c=1$ Matrix Model performed in \cite{srd}, and more recently revisited and improved in \cite{hartnoll}, clearly exhibits how the coupling constant plays the role of a UV cutoff, as expected in a theory of gravity.

The 1/2 BPS states of $\mathcal{N}=4$ SYM are known to represent giant gravitons and their duals \cite{giants} in supergravity - the fully backreacted geometries are the
well-known LLM geometries \cite{llm}. This duality has been explored
from various points of view \cite{mandal1, rychkov, others, skenderis}. 
In this context the Thomas-Fermi approximation is particularly relevant. Filled regions of the phase space are in one-to-one correspondence to the classical solutions of supergravity \cite{llm}: the connection can be understood in terms of 
the collective coordinate quantization of LLM geometries \cite{mandal1}. 
In \cite{dmt1,dmt2} it was proposed
that target space entanglement is related to notions of entanglement
in the bulk. The current setup may provide a useful arena for exploring such a connection.

As we discuss below, LLL states can be described in terms of an auxiliary one dimensional system. The two point function of fermions in LLL states
\ben
 C(x_1,x_2;x_1',x_2') = \langle LLL|\psi^\dagger (x_1,x_2,t) \psi (x_1^\prime, x_2^\prime,t) | LLL\rangle 
 \label{4-defc}
 \een
can be then expressed as an integral over the one dimensional phase space density $\langle u(x,p,t)\rangle$ and the results of \cite{klich}-\cite{dhl1} can be then used to calculate the entanglement entropy in terms of $\langle u(x,p,t)\rangle$. As mentioned above, this becomes particularly simple in the Thomas-Fermi limit. We will choose the auxiliary system to be a harmonic oscillator so that our considerations directly apply to the holomorphic sector of the two dimensional isotropic oscillator. We argue that the classical value of the one dimensional phase space density provides a good approximation to the correlator (\ref{4-defc}) for $\sqrt{N}\ell \gg x_i,x_i^\prime \gg \ell$ where $\ell$ is the sole length scale of the problem (i.e. the magnetic length). In this regime we show agreement with a direct calculation in $2+1$ dimensions.

We perform an analytic calculation of entanglement entropy of  $N$ free fermions in a Lowest Landau Level state in this approximation.
Our strategy easily yields explicit expressions for {\em arbitrary} two dimensional subregions, and our approximation is adequate for subregions which are much larger than the underlying length scale of the theory but much smaller than the total area of the two dimensional plane \footnote{In this paper we will in fact work on an infinite plane}. The leading term is a perimeter law. We provide evidence that the coefficient of this perimeter law is independent of the shape of the subregion. We find that the coefficient we obtain is in excellent agreement with the calculations of \cite{sierra,corners}. In these papers, the entanglement entropy for specific subregions of samples of various shapes is calculated by directly evaluating the eigenvalues of the correlation matrix. The agreement provides evidence that the leading term in the cumulant expansion is a fairly good approximation.

However, when the entangling surface has corners, there are additional terms which depend only on angles at the corners and the fermion density \cite{corners} \footnote{Such additional contributions to the entanglement entropy are well studied for conformal field theories \cite{cornermyers}}. Our leading large-$N$ result for a subregion with perimeter ${\cal{P}} (\partial A)$ and $n_i$ corners with corner angle $\alpha_i$
\ben
S_A = \frac{\sqrt{\pi}}{6} \frac{{\cal{P}} (\partial A)}{\ell} +\sum_i n_i a (\alpha_i)
\label{00-1}
\een
where the length scale $\ell$ is related to the energy gap $\omega$ and the mass of the particle $m$ by $\ell^2 = 2/(m\omega)$ and the function $a(\alpha)$ is given by
\ben
a(\alpha) = - \frac{1}{12}\left[ 1 + 2  \frac{\cos \alpha}{|\sin \alpha|} \tan^{-1} \left|\cot \frac{\alpha}{2} \right|   \right]
\label{00-2}
\een
{Equations (\ref{00-1}) and (\ref{00-2}) are among our main results. 
In our treatment, the origin of these corner terms becomes transparent. The integrand involved in the expression for the entanglement entropy can be seen to involve the extrinsic curvature and its derivatives, and additional corner terms arise from singularities of these quantities. In fact the result (\ref{00-2}) is in exact agreement with the super-universal results for fluctuations proved in \cite{william2}. It is interesting that this result is exact in a Thomas-Fermi approximation. 

In \cite{corners} and \cite{sirois} the corner contributions were deduced by directly obtaining the entanglement spectrum for special entangling surfaces.
These results are in agreement with \cite{william2}, indicating  that the truncation to the lowest non-trivial cumulant is a good approximation for this quantity. The functional dependence of $a(\theta)$ remains unchanged if one takes into account higher order cumulants \cite{sirois}.}

{Finally, we discuss how this entanglement is related to a target space entanglement of the complex matrix model discussed above in specific states. Presently, we are able to make such a connection in a gauge fixed formalism as in \cite{dmt1}. A gauge invariant discussion along the lines of \cite{dmt2} should be possible. There could, however, be a subtlety in relating these results in a quadratic model to a target space entanglement of the 1/2 BPS sector of $\mathcal{N}=4$ SYM theory. Such an entanglement is associated with a subalgebra of operators obtained by a suitable projection: it remains to be seen if expectation values of these projected operators can be evaluated in the quadratic theory. If such an evaluation is possible, the one dimensional phase space density becomes really useful when one considers a class of excited states which are described by droplets and fermi surface deformations in phase space - these are giant gravitons or their duals, or Kaluza-Klein gravitons and the corresponding phase space density determines the supergravity solution. We will not pursue this latter aspect in this paper, but hope to report on results in a future communication.}

In section \ref{sec:two} we review the results of
\cite{klich,peschel,cala3} for the evaluation of the entanglement
entropy of a region $A$ in free fermion theory in terms of cumulants. We then express the leading contribution to integrals of
the phase space density, and evaluate these in the Thomas-Fermi approximation for a large number of fermions in a one dimensional potential, both in the ground state as well as in an excited state in a harmonic oscillator potential. In section \ref{sec:three} we review the connection of Lowest Landau Level states of fermions in an external magnetic field to those of an auxiliary one dimensional fermion theory.
Beginning in section \ref{sec:four} we choose the auxiliary system
as a one dimensional harmonic oscillator and derive general expressions for the fermion 2 point
function. In section \ref{halfbps} we review the connection of the
LLL problem with those of 1/2-BPS states of $\mathcal{N}=4$ theory. In section
\ref{sec:EE} we derive the general expression for the entanglement
entropy for arbitrary subregions in the large-$N$ limit. Section
\ref{sec:smooth} contains the calculation of
the entanglement entropy for smooth entangling curves. In section
\ref{sec:corner} we summarize the calculation for entangling curves
with sharp corners. Section \ref{sec:target} discusses the
relationship of the entanglement entropy computed in this paper with
target space entanglement. The discussion section
contains some remarks about the relevance of our
calculation to gauge-gravity duality. Details of the calculations are
given in the appendices.

\section{Entanglement entropy in terms of phase space density}
\label{sec:two}

For a system of $N$ free non-relativistic fermions, it has been known for a while that the entanglement entropy of a subregion $A$ can be expressed as a sum over cumulants of the particle number distribution \cite{klich}
\ben
S_A = {\rm lim}_{M \rightarrow \infty} \sum_{m=1}^{M} \alpha_{2m} (M) C_{2m}
\label{2-1}
\een
where the cumulant $C_m$ is defined as
\ben
C_m = (-i\partial_\lambda) \log \langle [ \exp(i\lambda \hN_A)] \rangle
\vert_{\lambda = 0}
\label{2-2}
\een
where $\hN_A$ is the particle number operator in the region $A$, 
\ben
\hN_A = \int_A d^d\vx~ \hpsi^\dagger (\vx) \hpsi (\vx)
\label{2-3}
\een
where $\hpsi (\vx)$ is the fermion field and $d$ is the dimensionality of space. The coefficients $\alpha_{2m}$ are pure numbers given in \cite{klich}. The expression (\ref{2-1}) is useful when a finite number of terms in the infinite series contribute. Indeed for several systems it is known that in the leading order of large-$N$ limit the second cumulant dominates \cite{cala3}, including the system of interest \cite{klichqhe}. This leading result is 
\ben
S_A^{leading} = \frac{\pi^2}{3} \left[ \langle \hN_A^2 \rangle - \langle \hN_A \rangle^2 \right]
\label{2-4}
\een
This expression has a rather simple form for an interesting class of states. Consider the expansion of the Schrodinger picture fermion field in terms of modes, which we assume to be discrete and labelled by a set of integers $\{ m_i \} \equiv M$
\ben
\hpsi (\vq) = \sum_M\hc_M \phi_M (\vq) 
\label{2-5}
\een
Let us divide the labels $M$ into two sets, $\{ M_1 \}$ and $\{M_2 \}$ and a state of the $N$ fermion state $|\Omega\rangle$ such that
\bea
\hc^\dagger_M |\Omega \rangle & = & 0~~~~~~~~ M \in \{ M_1 \} \nonumber \\
\hc_M |\Omega \rangle & = & 0~~~~~~~~ M \in \{ M_2 \} 
\label{2-6}
\eea
A simple calculation then shows that (\ref{2-4}) becomes \cite{cala3}
\ben
S_A^{leading} = \frac{\pi^2}{3} \left[ \langle \Omega | \hN_A | \Omega \rangle - \int_A d^d\vx \int_A d^d\vx^\prime \vert \langle \Omega | \hpsi^\dagger (\vx) \hpsi (\vx^\prime) |  \Omega \rangle \vert^2 \right]
\label{2-7}
\een
Examples of states like $|\Omega \rangle$ include the ground state where the lowest $N$ single particle levels are filled, or Bogoliubov transformations of such a state. 

The equation (\ref{2-7}) can now be re-written in terms of the phase space density operator
\begin{align}
\frac{3 S_A^{leading}}{\pi^2} & = \frac{1}{(2\pi \hbar)^{d}} \int d^d\vp \int_A d^d\vx \langle \Omega | \hu(\vp,\vx) | \Omega \rangle \nonumber \\
& - \int \frac{d^d\vp_1 d^d\vp_2}{(2\pi\hbar)^{2d}} \int_A d^d\vx_1 \int_A d^d\vx_2 ~e^{-i(\vp_2-\vp_1)\cdot (\vx_2-\vx_1)/\hbar} ~\langle \Omega | \hu\bigg(\vp_1, \frac{\vx_1+\vx_2}{2}\bigg)|  \Omega \rangle \langle \Omega | \hu\bigg(\vp_2, \frac{\vx_1+\vx_2}{2}\bigg)|  \Omega \rangle \nonumber \\ 
\label{2-9}
\end{align}
This form of the entanglement entropy is very useful for the following reason. In the limit 
\ben
N \rightarrow \infty, \hbar \rightarrow 0~~~~~N\hbar = {\rm finite}
\label{22-5}
\een
and in regions far from turning points where the local fermi momentum is a slowly varying function, one can use a Thomas-Fermi approximation, i.e. $\langle \Omega | u(p,x) | \Omega \rangle$ is either zero or unity (for a recent discussion of the validity of this approximation, see  e.g.\cite{satya3}). Thus, the leading expression for the entanglement entropy becomes expressible in terms of integrals of simple functions.

The higher order terms in the culumant expansion (\ref{2-2}) can be similarly expressed in terms of the two point function
\ben
C_A (\vx,\vx^\prime) = \langle \Omega | \psi^\dagger (\vx) \psi (\vx^\prime) | \Omega \rangle
\label{22-6}
\een
and therefore in terms of the expectation value of the phase space density. Regarding $C_A (\vx,\vx^\prime)$ as a matrix with $\vx,\vx^\prime$ as indices, these are expressible in terms of the matrix
\ben
E_A \equiv C_A (1-C_A)
\label{22-7}
\een
For example, the next non-trivial cumulant, $C_4$ is given by
\ben
C_4 = {\rm Tr} \left[ E_A - 6 E_A^2 \right]
\label{22-8}
\een

\subsection{One dimensional potentials}

Let us see this for the example of $N$ fermions in a one dimensional confining potential with the hamiltonian
\ben
H = \frac{1}{2m}\hp^2 +  V(x) 
\label{2-11}
\een

\subsubsection{Ground state}

In the semiclassical large $N$ limit (\ref{22-5}) the ground state expectation value of the phase space density is given by 
\ben
\langle \Omega| \hu(p,x) | \Omega \rangle = \Theta \bigg(2\mu_F - \frac{p^2}{2m}-V(x)\bigg)
\label{22-10}
\een
where $\mu_F$ is the fermi energy. To examine the regime in which this approximation holds, consider the two point function of the fermion field, 
\ben
\langle \Omega | \psi^\dagger (x_1) \psi (x_2) |\Omega \rangle = 
\int \frac{dp}{2\pi\hbar} e^{-ip(x_2-x_1)/\hbar} \langle \Omega | \hu \left( \frac{x_1+x_2}{2},p \right)|\Omega\rangle
\label{62-1}
\een
In the Thomas-Fermi approximation, equation (\ref{22-10}), this is given by
\ben
\langle \Omega | \psi^\dagger (x_1) \psi (x_2) |\Omega \rangle
= \frac{1}{\pi (x_2-x_1)}\sin \left[ \frac{(x_2-x_1)}{\hbar} P_F \left( \frac{x_1+x_2}{2}\right) \right]
\label{a1-1-1}
\een
where $P_F(x)$ denotes the local fermi momentum,
\ben
P_F(x) = \sqrt{2m(\mu_F - V(x))}
\label{62-5}
\een
As shown in equation (14) of \cite{satya2} this is the correlator in the microscopic limit $|x_2-x_1| \sim \hbar/P_F(x)$. In addition, both the points $x_1$ and $x_2$ must lie in the regime of validity of the WKB approximation
\ben
\vert \frac{\hbar}{(P_F(x))^2}\frac{d P_F(x)}{dx} \vert \ll 1
\label{62-6}
\een
The fermi energy $\mu_F$ is proportional to $N$. Note that for $N \gg 1$ there is typically a large region of $x$ away from the turning point over which (\ref{62-6}) holds. Consider for example the harmonic oscillator potential $V(x) = \frac{1}{2}m\omega^2 x^2$. The fermi energy is then $\mu_F = N \hbar \omega$. In terms of the usual dimensionless coordinate
\ben
y = x \sqrt{\frac{m\omega}{\hbar}}
\label{62-7}
\een
the condition (\ref{62-6}) reads
\ben
y \ll (2N - y^2)^{3/2}
\een
For $N \gg 1$ there is a fairly large regime of $y$ over which this holds.

Consider the entanglement entropy of a small interval $x_0-a < x < x_0+a$ which lies in this regime. Since the fermi momentum $P_F(x)$ is slowly varying, to lowest order we can treat this to be constant and the integrals in (\ref{2-7}) can be performed analytically. This leads to the result
\bea
S &=& \frac{\pi}{3\hbar} ~ 2a P_F(x_0) - \frac{1}{3} \left[
\frac{4 P_F(x_0)}{\hbar} \text{Si}(4P_F(x_0)a/\hbar)+ \text{Ci}(4P_F(x_0)a/\hbar) + \cos (4P_F(x_0)a/\hbar) \right] \nonumber \\
&&  + \frac{1}{3}[1+\gamma_E +\log(4P_F(x_0)a/\hbar)]
\label{62-8}
\eea
where $\text{Si}(x)$ and $\text{Ci}(x)$ are CosIntegral and SinIntegral functions respectively.When the interval lies in the regime
\ben
4P_F(x_0)a/\hbar \gg 1
\label{62-9}
\een
we can use the asymptotic expansions of $\text{Si}(x)$ and $\text{Ci}(x)$,
\ben
\text{Si}(x) \sim \frac{\pi}{2} - \frac{\cos(x)}{x} + O(1/x^2)~~~~~~
\text{Ci}(x) \sim \frac{\sin(x)}{x} + O(1/x^2)
\label{62-10}
\een
 The terms linear in $a$ cancel, and we get the result
\ben
S_{EE} = \frac{1}{3}\left[ 1 + \gamma_E + \log \left(4P_F(x_0)a/\hbar\right) \right]
\label{2-13}
\een
For a harmonic oscillator potential this should be compared with equation (1) of \cite{satya2}.

\subsubsection{Excited states in 1d harmonic oscillator}\label{2-1-2}

The expression of the large $N$ entanglement entropy in terms of the phase space density becomes really useful for a class of excited states which are described easily in terms of filled regions in phase space. To illustrate this, let us consider the one dimensional harmonic oscillator with $V(x) = \frac{1}{2}mx^2$. In the following we will set $m=1$, which can be easily restored.

Consider for example a classical phase space density
\bea\label{2-14}
\langle E | u(x,p) |E \rangle = \Theta(\sqrt{p^2+\omega^2 x^2}-r_1)\Theta(r_2-\sqrt{p^2+\omega^2 x^2})
\eea
where $|E\rangle$ represents an excited state.
The filled region in phase space has the geometry of an annulus of inner radius $r_1$ and width $d$ where
\bea\label{dp}
d&=&\sqrt{r_1^2 + 2N\hbar\omega}-r_1
\eea
which comes from the fact that the area of the annulus is $2N\hbar$. Figure \ref{annulus} shows the filled region in phase space.

\begin{figure}[H]
\centering
\includegraphics[width=0.4\textwidth]{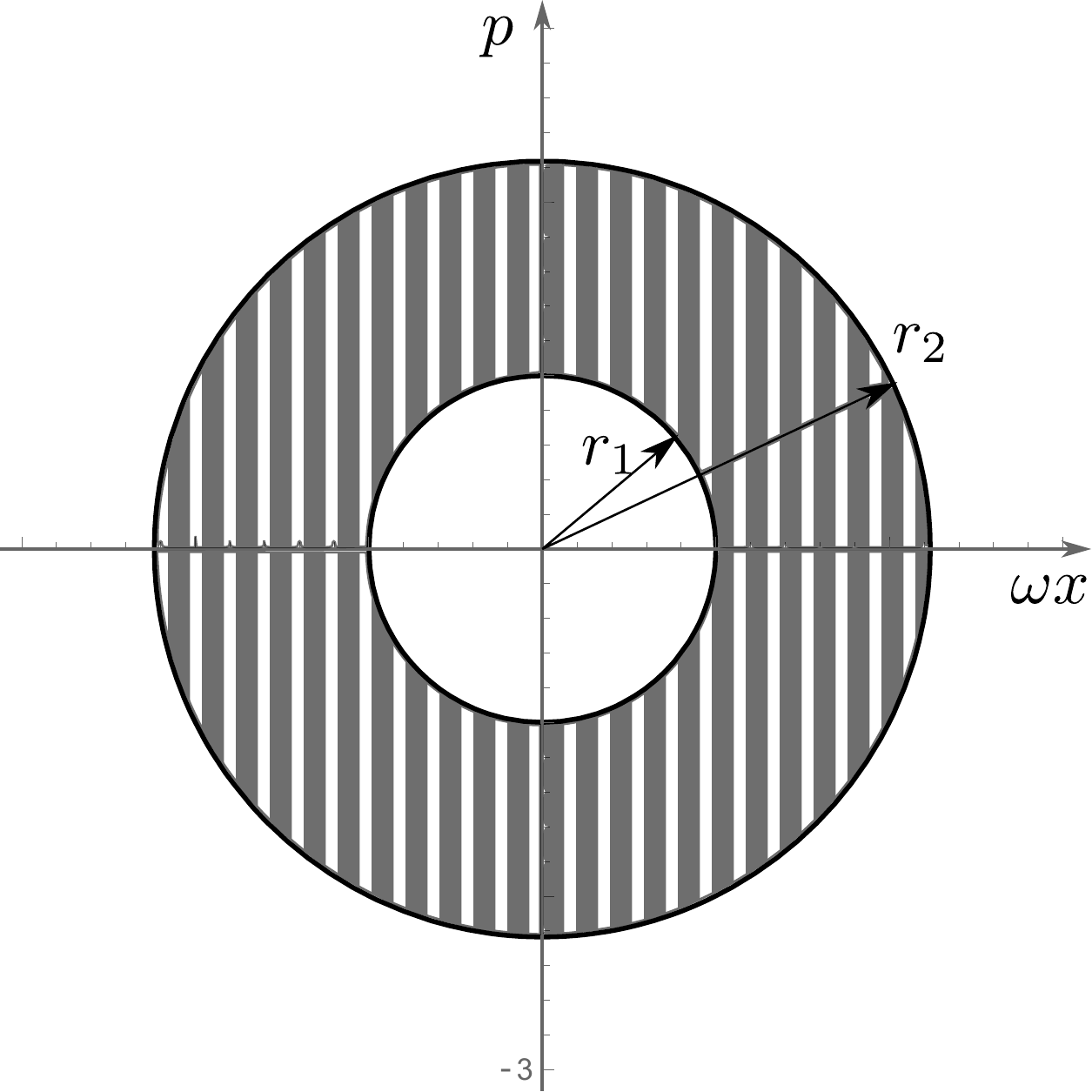}
\caption{Phase space picture of excited state described by (\ref{2-14})}
\label{annulus}
\end{figure}

In terms of the states of the harmonic oscillator this excited state has fermions in a set of contigious single particle states with energies between $r_1^2/2$ and $r_2^2/2$. Physically the value of $r_1$ describes how excited the state is.

Now consider a spatial interval $A$ of size $2a$ centered around a point $x_0$. When the interval is entirely within the annulus,
\bea\label{region}
A=\bigg\{ r_1/\omega + x_0 - {a}\leq x\leq r_1/\omega + x_0 + {a}\bigg\}
\eea 
a constant $x$ line cuts the edge of the filled annulus at values of the momenta 
$\pm P_0(x)$, where
\ben
P_0(x) = \sqrt{r_2^2 - \omega^2 x^2}
\label{62-11}
\een
The fermion correlator  (\ref{62-1}) with $r_1/\omega \leq x_1,x_2 \leq r_2/\omega$ in this approximation is given by equation (\ref{a1-1-1}) with $P_F(x_0)$ replaced by $P_0(x_0)$ above. The result for the entanglement entropy is exactly of the form (\ref{62-8}) with $P_F(x_0)$ replaced by $P_0(r_1+x_0)$. This is shown in appendix \ref{apponedim}.

When the interval lies in the region $-r_1/\omega \leq x \leq r_1/\omega$ a constant $x$ line intersects the edges of the annulus four times, at $\pm P_0(x), \pm P_1(x)$ where 
\ben
P_1(x) = \sqrt{r_1^2 - \omega^2 x^2}
\label{62-12}
\een
In the Thomas Fermi approximation the fermion correlator is now given by
\ben
\langle E | \psi^\dagger (x_1) \psi (x_2) |E \rangle
= \frac{1}{\pi (x_2-x_1)}\{ \sin \left[ P_0(x_c)(x_2-x_1)/\hbar \right] - 
\sin \left[(P_1(x_c)(x_2-x_1)/\hbar \right] \}
\label{62-13}
\een
where $x_c = \frac{x_1+x_2}{2}$.

In our approximation the expression for the entanglement entropy is
\bea
S_{EE} = \frac{2\pi a}{3\hbar} & & \left[P_0(x_0) -P_1(x_0) \right] \nonumber \\
& & -\frac{1}{3}\int_{x_0-a}^{x_0+a} \frac{dx_1 dx_2}{(x_2-x_1)^2}
\{ \sin \left[ P_0(x_c)(x_2-x_1)/\hbar \right] - 
\sin \left[(P_1(x_c)(x_2-x_1)/\hbar \right] \}^2 \nonumber \\
\label{62-14}
\eea
In performing these integrals $P_0(x_c),P_1(x_c)$ can be approximated by $P_0(x_0),P_1(x_0)$, since our approximation is valid only when the local momentum is slowly varying. In the limit where
\begin{equation}
\begin{split}
aP_0(x_0)/\hbar&\gg1\cr
aP_1(x_0)/\hbar&\gg1\cr
a(P_0(x_0)+P_1(x_0))/\hbar&\gg1\cr
a(P_0(x_0)-P_1(x_0))/\hbar&\gg1
\end{split}
\end{equation}
we find 
\bea\label{2-30}
S_{EE} = {1\over3}\bigg[2 + 2\gamma_E + \log\bigg({16a^2P_0(x_0)P_1(x_0)\over\hbar^2}\bigg({P_0(x_0) - P_1(x_0)\over P_0(x_0) + P_1(x_0)}\bigg)^2\bigg)\bigg]
\eea
This result is derived in appendix \ref{apponedim}. In the remaining part of the paper we will set $\hbar = 1$.

\section{Fermions in the Lowest Landau Level and auxiliary one dimensional system}
\label{sec:three}

The best known example of a Lowest Landau Level problem concerns free fermions with charge $-e$ moving on an infinite two dimensional plane with a constant magnetic field normal to it. This is the problem of Integer Quantum Hall Effect. The single particle hamiltonian is given by
\ben
H = \frac{(\hat{\vec{p}} + e \hat{\vec{A}}(\hat{\vec{x}}))^2}{2m}
\label{3-1}
\een
We will work in a symmetric gauge where the components of the vector potential are
\ben
\hA_1 = -\frac{B\hx_2}{2}~~~~~~~\hA_2 = \frac{B\hx_1}{2}
\label{3-2}
\een
and work on the infinite plane.
In terms of the operators
\bea
\ha  & =  & \frac{1}{\sqrt{2eB}} \left[ ({\hat{p}}_1 - \frac{eB\hx_2}{2})-i({\hat{p}}_2 + \frac{eB\hx_1}{2}) \right] \nonumber \\
\ha^\dagger  & =  &\frac{1}{\sqrt{2eB}} \left[ ({\hat{p}}_1 - \frac{eB\hx_2}{2}) + i({\hat{p}}_2 + \frac{eB\hx_1}{2}) \right]
\label{3-3}
\eea
The hamiltonian then becomes
\ben
H = \omega_B \left[\ha^\dagger \ha + \frac{1}{2} \right]~~~~~~~\omega_B = \frac{eB}{m}
\label{3-4}
\een
The basic length scale in the theory is given by \footnote{ Note that this differs from the standard definition of the magnetic length by a factor of $\sqrt{2}$}
\ben
\ell = \sqrt{\frac{2}{eB}}
\label{3-4a}
\een
In this gauge there is another set of oscillators which commute with the $\ha, \ha^\dagger$, defined by \cite{tong}
\bea
\hb & = & \frac{1}{\sqrt{2eB}} \left[ \bigg({\hat{p}}_2 - \frac{eB\hx_1}{2}\bigg)-i\bigg({\hat{p}}_1 + \frac{eB\hx_2}{2}\bigg) \right] \nonumber \\
\hb^\dagger & = & \frac{1}{\sqrt{2eB}} \left[ \bigg({\hat{p}}_1 - \frac{eB\hx_2}{2}\bigg) + i\bigg({\hat{p}}_2 + \frac{eB\hx_1}{2}\bigg) \right]
\label{3-5}
\eea
These obey $[ \hb , \hb^\dagger ] = 1$. The corresponding wavefunctions will be denoted by $\phi_{lm}(x_1,x_2)$

Therefore the states of the system can be constructed by the action of $\ha^\dagger$ and $\hb^\dagger$ acting on the Fock vacuum,
\bea
|m,n \rangle  & =  & \frac{1}{\sqrt{n !}\sqrt{m !}} (\ha^\dagger)^n (\hb^\dagger)^m |0 \rangle~~~~~~~~~\ha |0\rangle = \hb |0\rangle = 0 \nonumber \\
H |m,n \rangle & = & \bigg(n+\frac{1}{2}\bigg) |m,n \rangle
\label{3-6}
\eea
The quantum number $n$ labels a Landau level. 

Since the energy does not depend on $m$, there is an infinite degeneracy. This is because we have been working on an infinite plane. If we instead work on a plane with a finite but large area $A_2$, the degeneracy $d(n)$ is given by
\ben
d(n) = \alpha \frac{A_2}{\ell^2}
\label{3-7}
\een
where $\alpha$ is a numerical constant.

The Lowest Landau Level (LLL) states have $n=0$. The normalized wavefunctions are given by
\ben
\phi_{l,0}(x_1,x_2) = \frac{1}{\ell \sqrt{\pi l !}} \left( \frac{x_1+ix_2}{\ell} \right)^l~{\rm exp} \left( - \frac{1}{2\ell^2} [ x_1^2 + x_2^2] \right)
\label{3-8}
\een

Consider the second quantized fermion field in the two dimensional problem of section \ref{sec:two}
\ben
\hchi (x_1,x_2) = \hpsi (x_1,x_2) + \hlambda (x_1,x_2)
\label{4-1}
\een
where the first field operator is a sum over all the Lowest Landau Levels, while the second field operator is a sum over all the remaining higher Landau levels, i.e.
\bea
\hpsi (x_1,x_2) & = & \sum_{l=0}^\infty \hc_{l,0} \phi_{l,0} (x_1,x_2) \nonumber \\
\hlambda (x_1,x_2) & = &  \sum_{m=1}^\infty \sum_{l=0}^\infty \hc_{l,m} \phi_{l,m} (x_1,x_2)
\label{4-2}
\eea
The operators $\hc_{l,m}$ obey the anti-commutaton relations
\ben
\{ \hc^\dagger_{l,m}, \hc_{l^\prime,m^\prime} \} = \delta_{m m^\prime} \delta_{l l^\prime}~~~~~~~\{ \hc_{l,m}, \hc_{l^\prime,m^\prime} \} =0
\label{4-3}
\een
Since the full set of modes $\phi_{l,m}(x_1,x_2)$ form a complete set, 
the fermion field $\hchi (x_1,x_2)$ satisfies the standard anticommutation relations
\ben
\{ \hchi^\dagger (x_1,x_2), \hchi (x_1^\prime,x_2^\prime) \} = \delta(x_1-x_1^\prime)\delta (x_2-x_2^\prime)~~~~~~\{ \hchi (x_1,x_2), \hchi (x_1^\prime,x_2^\prime) \} = 0
\label{4-4}
\een
Of course, the operator $\hpsi (x_1,x_2) $ do not obey the standard anticommutation relations.

Let us now {\em define} a one dimensional fermion field operator by 
\ben
\hxi (v) \equiv \sum_{l=0}^\infty \hc_{l,0} \zeta_l (v)
\label{4-5}
\een
with the condition
\ben
\int_{-\infty}^\infty dv \hxi^\dagger (v) \hxi (v) = N
\label{4-5a}
\een
Here the functions $\zeta_l (v)$ form {\em any} complete orthonormal set on the real line,
\bea
\int_{-\infty}^\infty dv \zeta^*_l (v) \zeta_{l^\prime} (v) & = & \delta_{l l^\prime} \nonumber \\
\sum_{l=0}^\infty \zeta^*_l (v) \zeta_l (v^\prime) & = & \delta (v-v^\prime)
\label{4-6}
\eea
which ensures that 
\ben
\{ \hxi^\dagger (v), \hxi (v^\prime) \} = \delta (v-v^\prime)~~~~~~
\{ \hxi (v), \hxi (v^\prime) \} = 0
\label{4-6a}
\een

From the first equation in (\ref{4-2}), and the expression for $\hc_{l,0}$ in terms of $\hxi (v)$ obtained by inverting (\ref{4-5}) we get a relationship between the operator $\hpsi (x_1,x_2)$ and the one dimensional fermion field $\hxi(v)$,
\ben
\hpsi (x_1,x_2) = \int_{-\infty}^\infty dv~\calk (x_1,x_2;v) \hxi (v)
\label{4-7}
\een
where
\ben
\calk (x_1,x_2;v) = \sum_{l=0}^\infty \phi_{l,0} (x_1,x_2) \zeta^\star_l (v)
\label{4-8}
\een
We can regard the operator $\hxi (v)$ as the second quantized field for a set of $N$ free fermions living on the real line in the presence of a potential such that the modes $\zeta_l (v)$ are eigenfunctions of the corresponding hamiltonian. We will call this system a one dimensional auxiliary system. 
The operator $\hpsi (x_1,x_2)$ is a constrained field operator which satisfies the Lowest Landau Level condition at the operator level. The equation (\ref{4-7}) expresses this constrained operator in terms of an unconstrained one dimensional fermion field.

Finally the expression (\ref{4-8}) can be used to relate bilocals of the constrained two dimensional field to Wigner operators of the auxiliary one dimensional problem,
\bea
\hpsi^\dagger (x_1,x_2) \hpsi (x_1^\prime, x_2^\prime) 
& = & \int dv dv^\prime~\calk^\star (x_1,x_2;v) \calk (x_1^\prime, x_2^\prime;v^\prime) \hxi^\dagger (v) \hxi (v^\prime) \nonumber \\
& = & \int dv dv^\prime~\calk^\star (x_1,x_2;v) \calk (x_1^\prime, x_2^\prime;v^\prime) \int \frac{dp}{2\pi}~e^{-i\frac{v^\prime - v}{2}}~
\hu \bigg(p, \frac{1}{2} (v+v^\prime)\bigg) \nonumber 
\\ & & 
\label{4-9}
\eea
where the Wigner function of the auxiliary one dimensional problem is defined as in (\ref{2-8}),
\ben
\hu(p,v) = \int dv^\prime~\hxi^\dagger \bigg(v-\frac{v^\prime}{2}\bigg)\hxi \bigg(v + \frac{v^\prime}{2}\bigg)~e^{ipv^\prime} 
\label{4-10}
\een
Note that the modes $\zeta_l (v)$ can be {\em any set of complete orthonormal modes} so that the potential of the auxiliary one dimensional problem is arbitrary so long as the spectrum is discrete and labelled by an integer. Different auxiliary systems will have different correlators of $\hu(p,v)$ as well as different kernels $\calk(x_1,x_2;v)$ but would lead to the same correlators of $\psi(x_1,x_2)$.

\subsection{One dimensional harmonic oscillator as the auxiliary system}\label{sec:four}

In this paper we will choose the auxiliary one dimensional system to be a one dimensional harmonic oscillator so that the modes $\zeta_l(v)$ are the usual eigenfunctions expressible in terms of Hermite polynomials. As mentioned above this choice makes a direct connection to the LLL which appear as descriptions of the holomorphic sector of a {\em two} dimensional isotropic oscillator which is relevant to the 1/2 BPS problem.

We will be interested in the state where the $N$ fermions successively fill up the Lowest Landau Level single particle states,
\ben
|F\rangle \equiv \prod_{m=0}^N \frac{1}{\sqrt{m !}} (\hb^\dagger)^m |0\rangle
\label{4-12}
\een
This is the fermi ground state of the auxiliary one dimensional harmonic oscillator whose lowering and raising operators are given by the $\hb,\hb^\dagger$. The modes $\zeta_l(y)$ in (\ref{4-5}) are then given by
\ben
\zeta_l(v) = \frac{1}{\sqrt{2^l l! \sqrt{\pi}}}e^{-\frac{v^2}{2}}H_l(v)
\label{4-13}
\een
Note that the coordinate $v$ we use here is dimensionless. We can think of this as the usual dimensionless coordinate in the standard one dimensional harmonic oscillator. However, for us all we need are the normalization and completeness conditions (\ref{4-6}). The kernel $\calk (x_1,x_2;v)$ can be then easily evaluated (details are in appendix \ref{appkernel})
\ben
\calk (x_1,x_2;v) = \frac{1}{\pi^{3/4}\ell} {\rm exp} \left[ - \frac{1}{2\ell^2} ( x_1^2 +2i x_1x_2) - \frac{v^2}{2} + \frac{\sqrt{2} v}{\ell}(x_1 + ix_2) \right]
\label{4-14}
\een 
This can then be used in (\ref{4-9}) to obtain an explicit expression for the (two-dimensional) fermion bilinear in terms of the one dimensional phase space density. 

Consider now states for which the 1d phase space density $\langle u(p,v) \rangle$ is a function of $(p^2 + v^2)$ only. The expression for the two point function of fermions (\ref{4-defc}) can now be calculated using (\ref{4-14}) in (\ref{4-9}). To perform the integral we change variables to $(v \pm v^\prime)$ and the integral over $(v - v^\prime)$ can be performed explicitly. Making a further change of variables
\ben
(p, v + v^\prime) \rightarrow \bigg( r = \sqrt{p^2 + \frac{(v+v^\prime)^2}{4}},~~~ \tan \phi = \frac{2p}{(v+v^\prime)}\bigg)
\een
and utilizing the aforementioned symmetry of $\langle u(p,v) \rangle$ we get the expression
\begin{equation}
\begin{split}
C(x_1,x_2;x_1',x_2') 
= & \frac{2}{\pi \ell^2}  \exp \left[- \frac{x_1^2+x_2^2}{2 \ell^2} - \frac{x_1'^2+x_2'^2}{2 \ell^2} -2 \left( \frac{x_1-i x_2}{\sqrt{2} \ell} \right) \left( \frac{x_1'+i x_2'}{\sqrt{2} \ell} \right) \right] \\ 
& \times \int_{0}^{\infty}r dr ~e^{-r^2}\langle\chi| u (r^2)|\chi\rangle\cr
& \quad  I_0 \left(  2r \left[ 2\left( \frac{x_1-i x_2}{ \ell} \right)\left( \frac{ x_1'+i x_2'}{ \ell} \right) \right]^{1/2} \right) \\
\end{split}
\label{4-19}
\end{equation}
where $I_0(x)$ is the Modified Bessel Function. Notice that in the regime where
\begin{equation}
1 \ll \left| \left( \frac{x_1-i x_2}{ \ell} \right)\left( \frac{ x_1'+i x_2'}{ \ell} \right) \right| \ll N
\label{regA}
\end{equation}
the modified Bessel function can be replaced by its asymptotic behavior with a large argument, i.e. $I_0 (z) \to \frac{e^z}{\sqrt{2\pi z}}$. Thus the integral over $r$ can be calculated using the method of steepest descent, the saddle point of which is 
\begin{equation}
r^* = \left[ 2\left( \frac{x_1-i x_2}{ \ell} \right)\left( \frac{ x_1'+i x_2'}{ \ell} \right) \right]^{1/2}.
\end{equation}
This means that the main contribution to the integral comes from the regime around $\left[ 2\left( \frac{x_1-i x_2}{ \ell} \right)\left( \frac{ x_1'+i x_2'}{ \ell} \right) \right]^{1/2}$. This is still far away from the edge $r \sim \sqrt{2N}$ in the regime (\ref{regA}). Therefore, the Thomas-Fermi approximation for $\langle\chi| u (r^2)|\chi\rangle$ is valid in this regime.

For the state with $\langle u\rangle$ given by 
\ben
\langle\chi|u(p,v)|\chi\rangle = \Theta \left(2N - (p^2+v^2) \right)
\label{4-16a}
\een
we get
\begin{equation}
\begin{split}
C(x_1,x_2;x_1',x_2') 
=&  \frac{1}{\pi \ell^2}  \exp \left[- \frac{x_1^2+x_2^2}{2 \ell^2} - \frac{x_1'^2+x_2'^2}{2 \ell^2} -2 \left( \frac{x_1-i x_2}{\sqrt{2} \ell} \right) \left( \frac{x_1'+i x_2'}{\sqrt{2} \ell} \right) \right] \\  
& \times \sum _{m=0}^{\infty }\frac {1}{m!} \left[ 2\left( \frac{x_1-i x_2}{ \ell} \right)\left( \frac{ x_1'+i x_2'}{ \ell} \right) \right]^m \left[ 1- Q ( m+1, 2N )\right]
\end{split}
\label{4-20}
\end{equation}
where
\begin{equation}
\nonumber
  Q(s,\lambda) =1-\frac{1}{\Gamma(s)} \int_0^{\lambda} t^{s-1}\,e^{-t}\,{\rm d}t .\,\!
\end{equation}
The expression (\ref{4-20}) has a smooth limit as $N \rightarrow \infty$.  In the large-$N$ limit we finally have
\begin{equation}
C(x_1,x_2;x_1',x_2')  
= \frac{1}{\pi \ell^2}  \exp \left[- \frac{(x_1-x_1')^2}{2 \ell^2} - \frac{(x_2-x_2')^2}{2 \ell^2}  + i\frac{ x_1 x_2' - x_2 x_1'}{\ell^2} \right]
\label{4-21}
\end{equation}

In the state under consideration, it is in fact possible to directly calculate the correlation function using the LLL wavefunctions (\ref{3-8}). 
\begin{equation}
\begin{split}
 C(x_1,x_2;x_1',x_2') =& 
\frac{1}{\pi \ell^2} \sum_{l=0}^{\infty} \frac{1}{l!}\left[ \left( \frac{x_1 -ix_2}{\ell} \right) \left( \frac{x_1'+i x_2'}{\ell} \right) \right]^l \exp\left( -\frac{x_1^2 + x_2^2 + x_1'^2 + x_2'^2}{2\ell^2} \right) \\
&-\frac{1}{\pi \ell^2} \sum_{l=N}^{\infty} \frac{1}{l!}\left[ \left( \frac{x_1 -ix_2}{\ell} \right) \left( \frac{x_1'+i x_2'}{\ell} \right) \right]^l \exp\left( -\frac{x_1^2 + x_2^2 + x_1'^2 + x_2'^2}{2\ell^2} \right)
\end{split}
\end{equation}
The first term is exactly (\ref{4-21}) and the second term is exponentially suppressed in the regime
\begin{equation}
\left|  \left( \frac{x_1-i x_2}{ \ell} \right)\left( \frac{ x_1'+i x_2'}{ \ell} \right) \right| \ll N
\end{equation}
when $N$ is large. The details of the estimates of the correction is given in appendix \ref{appcrlfn}.

(\ref{4-21}) is the expression we will use to calculate the entanglement entropy of a subregion in the $(x_1,x_2)$ plane in the next section. This leading order correlator obeys the property
\ben
C(x_1,x_2;x_1',x_2')  = \int dx_1'' dx_2'' ~C(x_1,x_2;x_1'',x_2'') C(x_1'',x_2'';x_1',x_2') 
\label{4-22}
\een

While we are mainly concerned with the Thomas-Fermi approximation for the one dimensional phase space density in this paper, the two dimensional fermion number density is related directly to the phase space density near the edge of the filled region where the Thomas-Fermi approximation fails.
Using (\ref{4-14}) and (\ref{4-9}) we obtain 
\ben
\langle LLL | \psi^\dagger (x_1,x_2) \psi (x_1,x_2) |LLL \rangle = \int \frac{dpdv}{2\pi \ell^2} \langle \chi| u (p,v) | \chi \rangle_N ~
e^{ - (v-\sqrt{2} \frac{x_1}{\ell})^2 - (p  +\sqrt{2} \frac{x_2}{\ell})^2 }
\label{4-15}
\een
Here $| \chi \rangle_N$ is the state of $N$ fermions in the 1d oscillator which corresponds to the particular LLL state. Suppose this state is such that the expectation value  $\langle \chi | u (p,v) | \chi \rangle_N$ obeys the scaling property
\ben
\langle \chi | u (p,v) | \chi \rangle_N = f\bigg(\frac{p}{\sqrt{N}},\frac{v}{\sqrt{N}}\bigg)
\label{4-16}
\een
Then by a change of variables the integral in (\ref{4-15}) becomes
\ben
\langle LLL | \psi^\dagger (x_1,x_2) \psi (x_1,x_2) |LLL \rangle = N\int \frac{dpdv}{2\pi \ell^2} f(p,v) ~
e^{- N (v-\frac{\sqrt{2}x_1}{\ell \sqrt{N}})^2 - N(p  +\frac{\sqrt{2}x_2}{\ell \sqrt{N}})^2}  
\label{4-17}
\een
In the limit of $N \rightarrow \infty$, with $x_i \sim \ell\sqrt{N}$, the gaussians become delta functions, leading to 
\bea
\langle LLL | \psi^\dagger (x_1,x_2) \psi (x_1,x_2) |LLL \rangle & \rightarrow &  \frac{1}{2\pi \ell^2} f\bigg(-  \frac{\sqrt{2}x_2}{\ell \sqrt{N}},  \frac{\sqrt{2}x_1}{\ell \sqrt{N}}\bigg) \nonumber \\
&=& \frac{1}{2\pi \ell^2} \langle \chi | u \bigg(-\frac{\sqrt{2}x_2}{\ell }, \frac{\sqrt{2}x_1}{\ell}\bigg) | \chi \rangle_N
\label{4-18}
\eea
We will not use this relationship in what follows. However this could be useful in a treatment which goes beyond the semiclassical approximation.

\section{1/2 BPS states of $\mathcal N=4$ Yang Mills theory}
\label{halfbps}

In this section we review the connection of lowest Landau Level states and the 1/2 BPS sector of $\mathcal N=4$ Yang-Mills theory on $S^3$. As shown in \cite{antal}, the problem reduces to the holomorphic sector of quantum mechanics of a single complex matrix, which can be then expressed in terms of two dimensional fermions with holomorphic wave functions. Below we will follow the treatment of \cite{tsuchiya}.

This theory has six adjoint scalar fields $\phi_1 \cdots \phi_6$. Half BPS operators  are of the form
\ben
\cO^{(J_1 \cdots J_p)} = \prod_{a=1}^p {\rm Tr} (Z^{J_a})
\label{41-1}
\een
where $Z = \phi_1 + i\phi_2$ is a complex $N \times N$ matrix. The correlators of these operators can be computed in terms of the correlators of their lowest KK mode on $S^3$, i.e. we need to consider operators which depend only on time. Non-renormalization theorems then imply that the correlators of such operators can be calculated in terms of correlators of the singlet holomorphic sector of quantum mechanics of a single complex matrix with a hamiltonian
\ben
H = \sum_{i,j=1}^N \left( - \frac{\partial^2}{\partial Z_{ij} \partial Z^\star_{ij}} + Z_{ij} Z^\star_{ij} \right)
\label{41-2}
\een
The potential term comes from the conformal coupling of the scalar to the Ricci scalar of $S^3$. 

One way to reduce the theory to that of free fermions is to use a Schur decomposition \cite{tsuchiya}. The complex matrix can be now written in terms of a unitary matrix $U$ and a lower triangular matrix $T$
\ben
Z = U T U^\dagger
\label{41-22}
\een
and in the singlet sector the degrees of freedom are the matrix elements of $T$, which are $N$ complex numbers $z_i = T_{ii}$ with $i = 1\cdots N$ and $T_{ij}, T^\star_{ij},$ with $i < j$. The $z_i$ are the eigenvalues of $Z$.
The change of variables from $Z$ to these leads to a jacobian which is the modulus squared of the van der Monde determinant $\Delta (z) = \prod_{i<j}(z_i - z_j)$. Using the standard procedure of absorbing this in the wavefunction we get a theory of $N$ fermions moving in two dimensions with coordinates $(w_1)_i = \sqrt{2}{\rm Re}(z_i)$ and $(w_2)_i = \sqrt{2}{\rm Im}(z_i)$ and their conjugate momenta $(q_1,q_2)$ in an isotropic harmonic oscillator potential with a hamiltonian
\ben
h = \frac{1}{2} \left[ q_1^2 + q_2^2+ w_1^2 + w_2^2 \right]
\label{41-4}
\een
where $[ q_i , w_j ] = i$. In addition there are $\frac{1}{2}N(N-1)$ bosons in the same potential.

Let us introduce the oscillators
\ben
c_1 = \frac{1}{2}\left[ (w_1 + i w_2)+i(q_1 + iq_2) \right]~~~~
c_2 = \frac{1}{2}\left[ (w_1 - i w_2)+i(q_1 - iq_2) \right]
\label{41-6}
\een 
and their hermitian conjugates. The eigenstates of $h$ can be then written in the form
\ben
| n, m \rangle = (c_1^\dagger)^n (c_2^\dagger)^m |0\rangle = 0
\label{41-7}
\een

The 1/2-BPS sector of the original theory is then the sector where all the bosons are in their ground state, while the $N$ fermions occupy single particle states with wavefunctions of the form
\ben
\chi_{l} = \sqrt{\frac{2^l}{\pi l !}}~z^l~e^{-z^\star z}
\label{41-5}
\een
Holomorphic states are $|n,0\rangle$. 
The operators (\ref{41-1}) can be expressed in terms of the $z_i$'s. Since the bosons are always in their ground state, correlators of these operators can be expressed entirely in terms of the fermionic field theory.

The operators $c_1,c_2$ are in one-to-one correspondence with the operators  $\ha,\hb$ of the LLL problem defined in (\ref{3-3}) and (\ref{3-5}). The coordinates $x_1,x_2$ of the particle in a magnetic field are related to the coordinates of the isotropic harmonic oscillator problem by
\ben
x_i = \ell w_i~~~~p_i = \frac{q_i}{\ell}~~~~i=1,2
\label{41-8}
\een
which leads to 
\ben
\ha = - ic_2, \ha^\dagger = ic_2^\dagger~~~~~~\hb = - c_1, \hb^\dagger = - c_1^\dagger
\label{41-9}
\een
Note that the hamiltonian (\ref{41-4}) of the 2d oscillator is {\em not} the hamiltonian (\ref{3-1}) of the particle in a magnetic field. In fact, as is well known, the latter problem can be re-written in terms of a two dimensional harmonic oscillator deformed by a term proportional to the angular momentum. The holomorphic states are nevertheless mapped to LLL states.

Equivalently, in a coherent state representation, one could consider the matrix  $Z$ and its canonically conjugate momentum and define the matrix operators \cite{antal}
\ben
A_{ij} = \frac{1}{2} \left[ Z_{ij}+ \frac{\partial}{\partial Z^\dagger_{ij}} \right]~~~~~~B_{ij} = \frac{1}{2} \left[ Z^\dagger_{ij} - \frac{\partial}{\partial Z_{ij}} \right]
\label{41-10}
\een
The 1/2 BPS sector is the reduction to the matrix hilbert space given by the $A, A^\dagger$ oscillators. The holomorphic wavefunctions are then expressed in terms of the eigenvalues of $A$.

\section{The entanglement entropy: leading large $N$ expression}
\label{sec:EE}

Consider now a subregion $A$ of the $(x_1,x_2)$ plane. The correlation matrix for this region is
\ben
C_A(x_1,x_2;x_1',x_2') \equiv  \mathbf{1}_{(x_1,x_2) \in A} \langle  LLL | \psi^{\dagger}(x_1,x_2) \psi (x_1',x_2') | LLL \rangle \mathbf{1}_{(x_1',x_2') \in A} 
\label{5-1}
\end{equation}
where $\mathbf{1}_{(x_1,x_2) \in A}$ is the indicator function
\begin{equation}
\nonumber
\mathbf{1}_{(x_1,x_2) \in A} \equiv \left\{ 
\begin{matrix}
1, & (x_1,x_2) \in A \\
0, & (x_1,x_2) \notin A \\
\end{matrix}
\right.
\label{5-2}
\end{equation}
For the successively filled LLL states the leading term in the cumulant expansion of the entanglement entropy is given by the expression
\ben
S_A = \frac{\pi^2}{3} {\rm tr}E_A = \frac{\pi^2}{3}{\rm tr}[(I -C_A) C_A]
\label{5-3}
\een
Where the quantity $C_A$ is considered as a matrix with indices $(x_1,x_2)$ and $I$ is the identity matrix in this space. Using (\ref{4-22}) we can express $C_A$ in terms of a product of two 2-point functions. Furthermore, using the identity
\ben
I - (\mathbf{1}_{(x_1,x_2) \in A})^2 = \mathbf{1}_{(x_1,x_2) \in {A^c}}
\een
where $A^c$ denotes the complement of $A$, we can express the matrix $E_A$ as
\ben
 E_A (x_1,x_2; x_1'x_2')=  \int dx_1'' dx_2''~\mathbf{1}_{(x_1,x_2) \in A} C(x_1,x_2; x_1'',x_2'')\mathbf{1}_{(x_1'',x_2'') \in {A^c}}
C(x_1'',x_2'';x_1',x_2') \mathbf{1}_{(x_1',x_2') \in A}\nonumber \\
\label{5-4a}
\een
The expression for its trace is
\ben
\operatorname{tr} E_A  =
\int dx_1 dx_2 dx'_1 dx'_2~ | C(x_1,x_2;x_1',x_2')|^2 \mathbf{1}_{(x_1,x_2) \in A} \mathbf{1}_{(x_1',x_2') \in {A^c}}
\label{5-4b}
\een
This is an integral of the modulus squared of the fermion correlator between a point in $A$ and a point in $A^c$. This makes it clear that the dominant result will be a perimeter law. Using (\ref{4-21}) and (\ref{5-3}) the leading order expression for the entanglement entropy is therefore
\ben
S_A^{leading} = \frac{1}{3}\int_A dy_1 dy_2 \int_{A^c}dy_1^\prime dy_2^\prime~~
{\rm exp} [ - (y_1 -y_1^\prime)^2] 
 - (y_2 -y_2^\prime)^2] 
\label{5-33}
\een
where we have used dimensionless variables
\ben
y_{1,2} = \frac{x_{1,2}}{\ell}
\label{5-8}
\een
In the following, it will be useful to rewrite $E_A$ as
\ben
E_A = 
\mathbf{1}_{z + \epsilon/2 \in A}\cdot \mathbf{1}_{z - \epsilon/2 \in A}\int  \frac{ i d\eta d\eta^*}{2\pi^2 \ell^4} ~ \mathbf{1}_{z+\eta \in A^c}  \exp \left[ -\frac{|\epsilon|^2/2  + 2|\eta|^2-z\epsilon^* -\epsilon^* \eta  +z^*\epsilon  + \epsilon \eta^*}{2\ell^2}  \right] 
\label{5-4}
\een
where we have defined the complex variables
\bea
z & = & \frac{1}{2} (x_1+x_1'+ix_2 + ix_2') \nonumber \\
\epsilon & = &  (x_1-x_1'+ix_2 - ix_2') \nonumber \\
\eta & = & x_1''+ix_2'' - \frac{1}{2} (x_1+x_2'+ix_2 + ix_2')
\label{5-5}
\eea
To obtain the leading order entanglement entropy for the region $A$ given by (\ref{5-3}) we need to set $\epsilon = 0$ and perform the integration over $\eta,\eta^*$. This can be done by expanding
\ben
\mathbf{1}_{z+\eta \in A^c} = \sum_{k=0}^{\infty} \frac{1}{k!} (-1)^k \left( \eta \partial_z + \eta^* \partial_z^* \right)^k \mathbf{1}_{z \in A^c}
\label{5-6}
\een
One finally gets
\ben
\operatorname{tr} E_A  =
 \int d^2 x ~\mathbf{1}_{ \vec{x}  \in A} \frac{1}{\pi \ell^2}   \exp \left( \frac{\ell^2}{4} \nabla_{x}^2 \right) \mathbf{1}_{\vec{x} \in A^c}
\label{5-7}
\een

\section{Smooth entangling curves}
\label{sec:smooth}

In this section we will evaluate the leading large $N$ expression for the entanglement entropy (\ref{5-7}) for a smooth entangling curve.

\subsection{Entanglement of half space}
\label{halfspace}

Let us first consider the easiest case: entanglement of half space. The region $A$ is then defined by $y_1 > 0$. 
We will consider the case where the entire sample is a rectangle with sides $(2L_1,2L_2)$ in the $(y_1,y_2)$ direction (in $\ell=1$ units) and finally perform the limit $L_1,L_2 \gg 1$. The limits of integration are then $-L_1 \leq y_1 \leq L_1$ and $-L_2 \leq y_2 \leq L_2$.
Using (\ref{5-33}) we have
\ben
S_{{\rm{half-space}}}= \frac{1}{3} \int_0^\infty dy_1 \int_{-\infty}^0 dy_1^\prime 
{\rm exp} [ - (y_1 -y_1^\prime)^2] 
\int_{-\infty}^\infty dy_2 \int_{-\infty}^\infty dy_2^\prime ~{\rm exp} [ - (y_2 -y_2^\prime)^2] 
\label{55-1}
\een
In the limit $L_1 \rightarrow \infty$ straightforward integration yields the final result
\ben
S_{{\rm{half-space}}}= 2L_2 \frac{\sqrt{\pi}}{6 \ell}
\label{55-4}
\een
The details of the calculation of the integrals is given in appendix \ref{apphalf}.
As expected the result is proportional to the perimeter of the entangling surface, which is $2L_2$. In (\ref{55-4}) we have restored $\ell$.

\subsection{Circular entangling curve}
\label{circle}
Consider now a subregion $A$ which is bounded by a circle,
\ben
r \equiv \sqrt{y_1^2 + y_2^2} = r_0
\label{5-9}
\een
In polar coordinates the indicator functions are simply theta functions
\ben
\mathbf{1}_{ \vec{x}  \in A} = \Theta (r_0 -r)~~~~~~\mathbf{1}_{\vec{x} \in A^c} = \Theta(r-r_0)
\label{5-10}
\een
The expression (\ref{5-7}) becomes, after expanding the exponential and using (\ref{5-10})
\ben
\operatorname{tr} E_A = \int_0^{2\pi} d \theta \int_0^{r_0} dr~r ~ \frac{1}{\pi} 
\sum_{l=0}^\infty \frac{1}{l ! 2^{2l}} (\nabla^2)^l ~\Theta(r-r_0)
\label{5-11}
\een
This is an infinite series of terms which are individually singular. However, as we will show, in the limit $r_0 \gg 1$ (i.e. when the physical radius of the circle is much larger than the scale $\ell$), this sum has a finite answer. The $l=0$ term in (\ref{5-11}) clearly does not contribute. In the remaining terms we can bring out one factor of the laplacian and use Stokes' theorem to get
\ben
\operatorname{tr} E_A=\frac{r_0}{4\pi}\int_0^{2\pi} d\theta  \sum_{l=0}^{\infty} 
\frac{1}{(l+1) ! 2^{2l}} \left[ \partial_r \nabla^{2l} \Theta(r-r_0) \right]_{r=r_0}
\label{5-12}
\een
The terms inside the square bracket is a sum of terms of the form
\ben
 \left[ \partial_r \nabla^{2l} \Theta(r-r_0) \right]_{r=r_0} = \sum_{n=1}^{2l} \frac{a_{n,l}}{r_0^{2l-n}}\left[ \partial_r^{n}\delta (r-r_0) \right]_{r=r_0}
 \label{5-13}
 \een
 where $a_{n,l}$ are numerical coefficients. In the $r_0 \rightarrow \infty$ limit only the $n=2l$ term contributes and we can replace this by
\ben
\left[ \partial_r \nabla^{2l} \Theta(r-r_0) \right]_{r=r_0} \rightarrow 
\left[ \partial_r^{2l} \delta (r-r_0) \right]_{r=r_0}
\label{5-14}
\een
One way to see this concretely is to use e.g. the representation of the Dirac delta function as a limit of a gaussian whose width $\sigma \rightarrow 0$ and express the quantity $\left[ \partial_r^{n}\delta (r-r_0) \right]_{r=r_0}$ as a series of terms with successive inverse powers of $\sigma$. For a given $\sigma$, terms with inverse powers of $r_0$ will be suppressed. This leading order term can be evaluated by using the fourier representation of the delta function,
\ben
\left[ \partial_r^{2m} \delta (r-r_0) \right]_{r=r_0} = \int_{-\infty}^\infty \frac{dp}{2\pi} (-p^2)^m
\label{5-15}
\een
Using this in (\ref{5-14}) and (\ref{5-12}) 
we get
\bea
\operatorname{tr} E_A & = & \frac{r_0}{4\pi}\int_0^{2\pi} d\theta 
\int_{-\infty}^\infty \frac{dp}{2\pi} \sum_{l=0}^\infty \frac{1}{(l+1)!} \bigg(-\frac{p^2}{4}\bigg)^l 
=- \frac{r_0}{4\pi}\int_0^{2\pi} d\theta \int_{-\infty}^\infty \frac{dp}{2\pi}  \frac{4}{p^2} \left[ {\rm exp}(-p^2/4) - 1 \right] \nonumber \\
& = & \frac{1}{2 \pi^{3/2}} \int_0^{2\pi} r_0 d\theta
\label{5-16}
\eea
The final integral over $\theta$ is of course trivial. However we have kept the answer in the form (\ref{5-16}) to emphasize the fact that the integral is over arc lengths comprising the circle. The final answer for the entanglement entropy for a single circle is then
\ben
S_A = \frac{\pi^2}{3} \operatorname{tr} E_A  = (2\pi r_0) \frac{\sqrt{\pi}}{6 \ell}
\label{5-16-1}
\een
where we have restored the appropriate power of $\ell$. The result is again proportional to the perimeter $(2\pi r_0)$ with the coefficient which is exactly that in (\ref{55-4}).

It should be emphasized that this result is to leading order in $r_0 \gg 1$. 

\subsection{General smooth entangling curve}
\label{smooth}

The results of the previous subsection can be now generalized to arbitrary entangling curves provided that the extrinsic curvature and its derivatives at all points are small in units of $\ell$. Consider such a smooth curve $C$ parametrized by a parameter $\sigma$. Now approximate the interval between $\sigma$ and $\sigma + d\sigma$ by a circular arc with a radius $r (\sigma) = 1/K(\sigma)$ where $K(\sigma)$ is the extrinsic curvature scalar at the point $\sigma$. One might expect that as long as $K (\sigma) \ll 1/\ell$ at all points on the curve, we can calculate the contribution to $\operatorname{tr} E_A$ from this interval using the results of the previous subsection. A more careful consideration, detailed in appendix \ref{general}, shows that $K (\sigma) \ll 1/\ell$ is not enough. Rather one needs $|\partial_\sigma K(\sigma) | \ll |K(\sigma) | \ll {1/ \ell}$.

Under these conditions the area of the subregion $A$ enclosed by the curve will be large compared to $\ell^2$ and the leading contribution to the entanglement entropy is given by
\ben
S_A^{leading} = \frac{\pi^2}{3}V_A^{(2)} = \frac{\sqrt{\pi}}{6} \frac{{\cal{P}} (\partial A)}{\ell}
\label{5-17}
\een 
where ${\cal{P}} (\partial A)$ is the perimeter of the curve enclosing $A$.
This is the leading result in the limit $N \rightarrow \infty$ and $\frac{{\cal{P}} (\partial A)}{\ell} \gg 1$. The perimeter law is expected and has been derived previously using numerical methods. 

Taking into account the fact that our definition of the length scale $\ell$ differs from the magnetic length by a factor of $\sqrt{2}$, the coefficient of the perimeter law is in good agreement with e.g. the numerical result of \cite{sierra,corners}, which is $0.203$ which should be compared to our result $\frac{\sqrt{\pi}}{6\sqrt{2}}=0.208$.

\section{Corner contributions to entanglement entropy}
\label{sec:corner}

The perimeter law is the only leading term when the entangling curve has an extrinsic curvature small compared to $\ell$. When the curve has sharp corners, there are additional $O(1)$ corrections to the expressions of the previous section - these are the corner contributions. 

In our formalism the origin of these corner terms is the following. The expression (\ref{5-7}) involves the laplacian acting on the indicator function. Consider a subregion $A$ which is bounded by a closed curve $\cC : f(x_1,x_2) - r_0 = 0$. The indicator function is then 
\ben
\mathbf{1}_{ \vec{x}  \in A} = \Theta \left( r_0 - f(x_1,x_2 ) \right)
\label{6-1}
\een
It then follows that 
\ben
\partial_\mu \mathbf{1}_{ \vec{x}  \in A}= - n_\mu (\partial_\alpha f \partial^\alpha f)^{1/2} \delta \left( r_0 - f(x_1,x_2 ) \right)
\label{6-2}
\een
where $n_\mu = ( \partial_\mu f) / (\partial_\alpha f \partial^\alpha f)^{1/2}$ is the normal vector to the curve $\cC$. Thus
\ben
\partial_\mu \mathbf{1}_{ \vec{x}  \in A} = n_\mu (n^\alpha \partial_\alpha \mathbf{1}_{ \vec{x}  \in A}) 
\label{6-3}
\een
This leads to 
\ben
\nabla^2 \mathbf{1}_{ \vec{x}  \in A} = K (n^\alpha \partial_\alpha \mathbf{1}_{ \vec{x}  \in A}) + (n^\alpha \partial_\alpha)^2 \mathbf{1}_{ \vec{x}  \in A}
\label{6-3}
\een
where $K = \nabla^\mu n_\mu$ \footnote{Note that we are working in flat space using cartesian coordinates so that $\nabla_\mu = \partial_\mu$}
is the scalar extrinsic curvature of $\cC$ at this point. The approximation used in subsections \ref{circle} and \ref{smooth} amounts
to the replacement  $\nabla^2 \rightarrow (n^\alpha \partial_\alpha)^2$ which is valid when the extrinsic curvature $K \ll 1$ in $\ell =1$ units.
At a sharp corner $K$ is large and this approximation clearly breaks down, leading to additional terms.

\subsection{Corner terms from a triangular region}

To isolate these corner terms, it is useful to consider a subregion which is an isosceles triangle. To isolate the corner term for an arbitrary curve with corners we can replace the region near the corner by such a triangle.
\ben
A : 0 \leq y_1 \leq h,~~-ky_1 \leq y_2 \leq ky_1
\label{6-3}
\een
Here the parameter $k$ is related to the angle at the vertex $\alpha$ by
\ben
k = \tan \frac{\alpha}{2}
\label{6-4}
\een
We will be interested in the case $h \gg 1$.

For this calculation it is convenient to use the form (\ref{5-3}). The first term on the right hand side of this equation is proportional the number of fermions in the subregion, $N_A$. The fermion density is of course constant so this leads to a contribution to the leading entanglement entropy which is proportional to the area,
\ben
S_1 = \frac{\pi}{3} h^2 k
\label{6-5}
\een
where we used (\ref{4-22}).
As we will see soon, this term will be cancelled by contributions coming from the second term in  (\ref{5-3}). The latter is
\ben
S_2 = - \frac{1}{3}\int_0^h dy_1 \int_0^h dy^\prime_1 \int_{-ky_1}^{ky_1}dy_2 \int_{-ky_1^\prime}^{ky_1^\prime}dx_2^\prime~{\rm exp} \left[-(y_1-y_1^\prime)^2 - (y_2 - y_2^\prime)^2 \right]
\label{6-6}
\een
The integrals can be performed for $h \gg 1$. The details are given in appendix \ref{apptriangle}. The final result is
\bea
S_2 &=& -\frac{\pi}{3} h^2 k + \frac{\sqrt{\pi}}{6} \left(2h(k+\sqrt{1+k^2})\right)- \frac{\pi^2}{3}\left[\frac{k}{2\pi}+\frac{3}{4\pi^2}+\frac{{\rm cot}^{-1}(k)}{4\pi^2} \bigg(\frac{1}{k}-3k \bigg) \right]\cr
 &&+ O(e^{-h^2}) + O(e^{-h^2k^2})
\label{6-7}
\eea
The second term in $S_2$ is proportional to the perimeter
\ben
\calp (k;h) = 2h(k+\sqrt{1+k^2})
\label{6-8}
\een
and the coefficient is exactly the same as in the previous cases, equations (\ref{55-4}, \ref{5-16-1}, \ref{5-17}). Adding (\ref{6-5}) and (\ref{6-8}) we get
\ben
S_A = \frac{\sqrt{\pi}}{6}\calp (\alpha;h) -\frac{1}{12 \sin \alpha} \left[ \pi + \alpha + 3 \sin \alpha - 2 \alpha \cos \alpha \right]+ O(e^{-h^2}) + O(e^{-h^2 \tan^2 \frac{\alpha}{2}})
\label{6-9}
\een
where we have expressed the result in terms of the vertex angle $\alpha$. This result is derived in appendix \ref{apptriangle}.

\begin{figure}[H]
\centering
\includegraphics[width=0.7\textwidth]{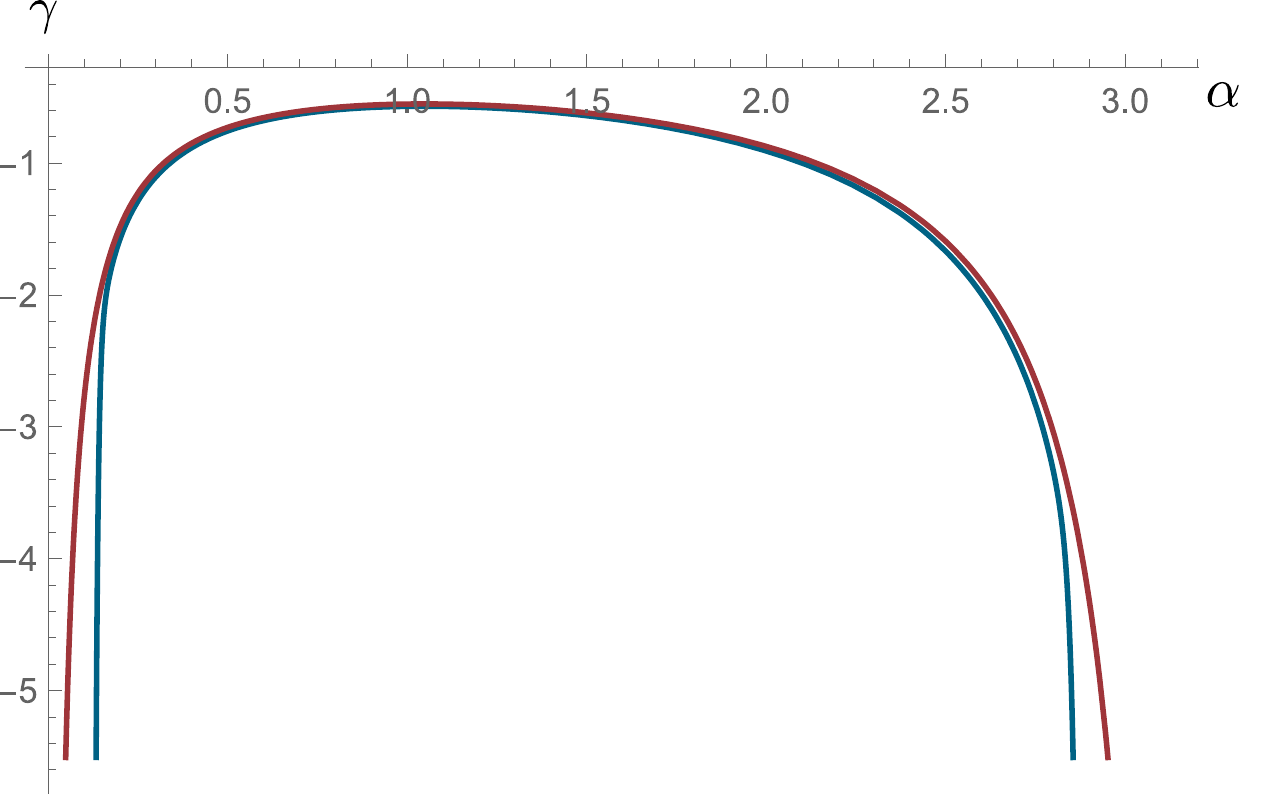}
\caption{EE for isosceles triangle : comparison between our results (in red) given by $S_A - \frac{\sqrt{\pi}}{6}\calp (\alpha;h)$ in equation (\ref{6-7}) and the prediction from numerical results (in blue) in \cite{corners}.}
\label{isotri}
\end{figure}

We now compare this result with that of \cite{corners}. In this paper the corner contributions are expressed as 
\ben
\gamma = \sum_{i=1}^p a(\alpha_i) n_i
\label{7-12}
\een
where a polygonal subregion has $n_i$ vertices with angles $\alpha_i$. The expression for $a(\alpha)$ obtained by numerical calculation for various different subregions is given in equation (22) of \cite{corners}. Figure \ref{isotri} compares our result for the corner contribution for an isosceles triangle, $S_A - \frac{\sqrt{\pi}}{6}\calp (\alpha; h)$ in equation (\ref{6-7}), the red curve, with the result obtained using equation (22) of \cite{corners}, the blue curve. We find a very good agreement away from $\alpha = 0$ or $\alpha = \pi$. However for these latter values of the vertex angle, the triangle degenerates and the approximations we have used to arrive at (\ref{6-7}) break down. 

\subsection{A pie-slice subregion and a better evaluation of the corner term}

To obtain a more general expression for the corner contribution we consider a subregion $A$ which is a pie-slice of the entire space. In terms of polar coordinates $(r,\phi)$ (with $y_1=r\cos \phi, y_2 = r\sin \phi$) this is given by
\ben
A: 0 \leq r \leq \infty~~-\frac{\alpha}{2} \leq \phi \leq \frac{\alpha}{2}
\label{8-1}
\een
We will, introduce a IR regulator which effectively provides an upper limit to the range of $r$. When $\alpha = \pi$ this is the calculation in subsection \ref{halfspace}. The leading order expression for the entanglement entropy is (using polar coordinates in the expression (\ref{5-33}))
\ben
S (\alpha,\epsilon) = \frac{1}{3} \int_{-\alpha/2}^{\alpha/2} d\phi_1
\int_{-\pi +\alpha/2}^{\pi -\alpha/2} d\phi_2 \int_0^\infty dr_1 dr_2~r_1r_2~ {\rm exp} [-(1+\epsilon)(r_1^2 + r_2^2) - 2r_1r_2 \cos (\phi_1 - \phi_2) ]
\label{8-2}
\een
Here we have introduced a small parameter $\epsilon$. Since the subregion (\ref{8-1}) is infinite the integral over $r$ would lead to a IR divergence. However this should be in fact the perimeter piece. The parameter $\epsilon$ effectively cuts off this integral to $r_{max} \sim 1/\sqrt{\epsilon}$. 

In fact, since we have already performed the integral (\ref{8-2}), we will compute the quantity
\ben
a(\alpha) = {\rm Lim}_{\epsilon \rightarrow 0} \left[ S (\alpha,\epsilon) -
S (\pi,\epsilon) \right]
\label{8-3}
\een
which is then identified with the corner contribution for a corner angle $\alpha$.

The integral (\ref{8-2}) is evaluated in appendix \ref{apppie}. The final result is
\bea
S(\alpha,\epsilon) & = & -\frac{1}{24}\left[  \log 2\epsilon  -3 -2 \operatorname{Li}_1 (1) \right]\left( 1 + 2 \operatorname{Li}_0 (1) \right) \nonumber \\
& & - \frac{1}{12}\left[ 1 + 2  \frac{\cos \alpha}{|\sin \alpha|} \tan^{-1} \left|\cot \frac{\alpha}{2} \right|   \right]
\label{8-4}
\eea
where $\operatorname{Li}_s(z)$ denotes the Polylogarithm function,
\ben
\operatorname{Li}_s(z) = \sum_{k=1}^\infty \frac{z^k}{k^s}
\een

The first line in (\ref{8-4}) is independent of $\alpha$ and is in fact formally divergent. The $\alpha$ dependence is entirely in the second line, which is independent of $\epsilon$, and vanishes at $\alpha = \pi$,
\ben
1 + 2  \frac{\cos \alpha}{|\sin \alpha|} \tan^{-1} \left|\cot \frac{\alpha}{2} \right| = \frac{1}{3}(\pi -\alpha)^2 +\mathcal{O}((\pi -\alpha)^4) 
\label{8-5}
\een
This means that the quantity $a(\alpha)$ is finite,
\ben
a(\alpha) = - \frac{1}{12}\left[ 1 + 2  \frac{\cos \alpha}{|\sin \alpha|} \tan^{-1} \left|\cot \frac{\alpha}{2} \right|   \right]
\label{8-6}
\een
This is our final result for the corner contribution. {This result is in {\em exact} agreement with the super-universal result for fluctuations derived in \cite{william2}. This agreement is an evidence for the validity of the Thomas-Fermi approximation for the expectation value of the phase space density.}

\begin{figure}
\centering
\includegraphics[width=0.75\textwidth]{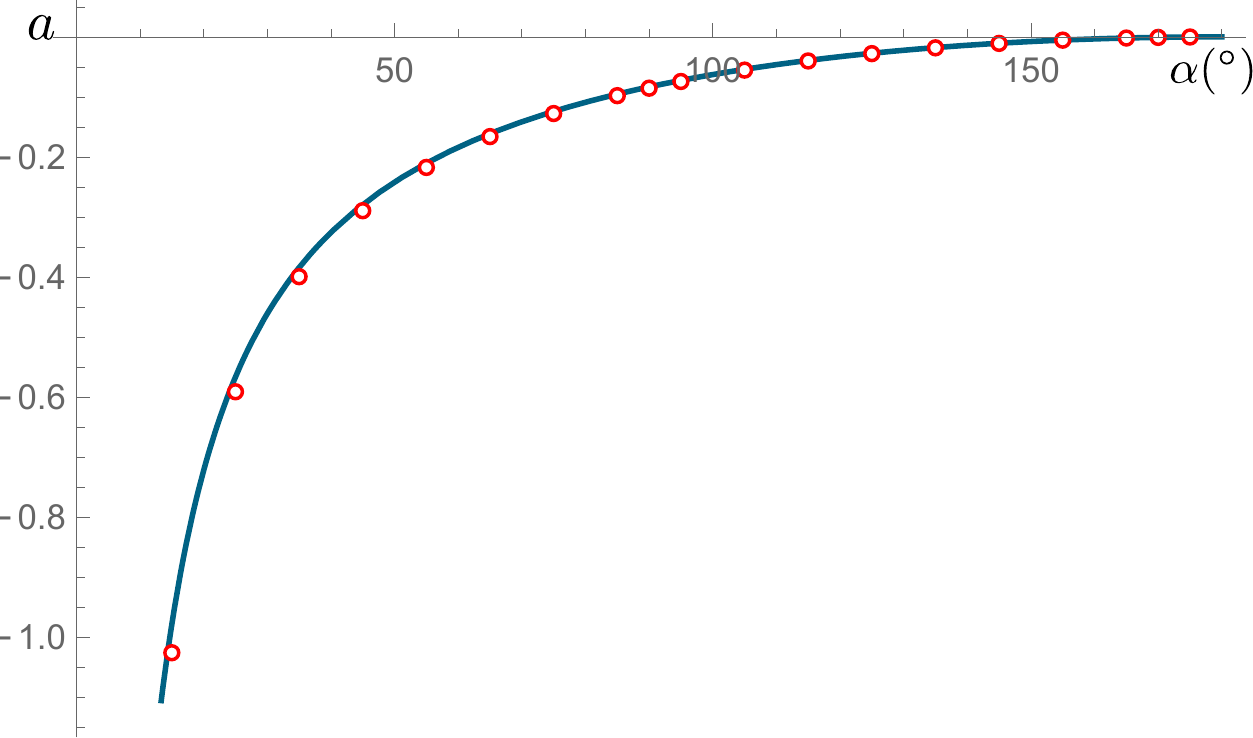}
\caption{Plot of the corner contribution to entanglement entropy from a single corner $a(\alpha)$ in equation (\ref{8-6}) (in blue solid line). This is compared with the results of \cite{sirois} which are red dots.}
\label{cornerplot}
\end{figure}

In \cite{corners,sirois} the corner contribution to the entanglement entropy has been numerically evaluated for special classes of entangling curves without truncation to the lowest cumulant, and expressions for $a(\alpha)$ have been deduced from these results. Our result is in good agreement with these results, as shown in figure \ref{cornerplot}. The difference is around $2-3 \%$, as tabulated in Appendix I. This shows that the truncation to the lowest cumulant is reasonably good.

\section{LLL entanglement as target space entanglement in $\mathcal{N}=4$ SYM}
\label{sec:target}

As explained in section \ref{halfbps} the coordinates of the fermions in the Lowest Landau Level are the eigenvalues of the complex matrix $Z$ of the $\mathcal N=4$ theory. We have used the second quantized formalism where the entanglement entropy we computed measures the entanglement among regions of the base space of this fermionic field theory. In this section we comment on the meaning of this quantity in the matrix model itself. 

The complex matrix model (\ref{41-2}) evaluates the correlators of operators which are of the form (\ref{41-1}). In section \ref{halfbps} we used a gauge in which the degrees of freedom are those in the lower triangular matrix $T$ defined in (\ref{41-22}). The diagonal elements $z_i = T_{ii} = (w_1+iw_2)_i$ are the eigenvalues of $Z_{ij}$. These, together with $R_{ij} \equiv T_{ij}, i < j$ form the target space of the quantum mechanical model. In this gauge the operator (\ref{41-2}) is
\ben
\cO^{(J_1 \cdots J_p)} = \prod_{a=1}^p \left[\sum_{i=1}^N( z_i^{J_p}) \right]
\label {9-1}
\een
The wavefunctions which are in one-to-one correspondence with states in the Lowest Landau level are of the form
\ben
\Psi(z_i,\bz_i, R_{ij}) = {\rm det} \left[ z_{i}^{l_j} \right] {\rm exp} \left[ -\sum_{i=1} (z_i^\star z_i) - \sum_{i<j}R^\star_{ij}R_{ij}\right]
\label{9-2}
\een
A basis in the Hilbert space is provided by the simultaneous eigenstates
$| \vw_i, R_{ij} \rangle$, where we have used the notation
$\vw = (w_1,w_2)$. Since all states in this sector have the same dependence on $R_{ij}$ and the operators (\ref{9-1}) do not involve the $R_{ij}$'s, the density matrix associated with a state of the form (\ref{9-2}) may be written as
\ben
\rho \{ \vw_i, R_{ij}; \vw^\prime, R^\prime_{ij} \} = \brho \{ \vw_i ; \vw^\prime_i \}~{\rm exp} \left[ - \sum_{i<j}(R^\prime_{ij})^\star R^\prime_{ij} \right]
\label{9-3}
\een
The expectation value of operators $\cO$ of the form (\ref{9-1}) are then given by ${\rm Tr} (\cO \brho)$.

A sub-region $A$ of the $(w_1,w_2)$ space (or equivalently $(z, z^\star)$ is a region of the target space, and the entanglement entropy we calculated is the target space entanglement entropy as discussed in \cite{dmt1, dmt2,mazenc, lawrence}. 
As discussed in these papers, the Hilbert space becomes a sum over sectors $(n, N-n)$ where $n$ denotes the number of eigenvalues which lie in the chosen region of the $(y_1,y_2)$ space. As discussed in \cite{dmt1,dmt2,mazenc}, the reduced density matrix in this sector is given by
\ben
\trho_n \{ \vw_a; \vw^\prime_a \} = \int_{A^c} \prod_\alpha d^2w_\alpha~\brho \{ \vw_a, \vw_\alpha ; \vw^\prime_a, \vw_\alpha \}
\label{9-4}
\een
where we have split the variables $(\vw)_i = \{ (\vw)_a, (\vw)_\alpha \}$
with $a=1 \cdots n, \alpha = n+1 \cdots N$ and the integral over the complement $A^c$ of the sub-region $A$. The target space entanglement entropy is then given by a sum over the sectors
\ben
S = - \sum_{n=0}^N \trho_n \log \trho_n
\label{9-5}
\een
As shown in \cite{dmt1} this is the same quantity which is computed in the second quantized formalism. 

The preceding discussion is in a gauge fixed setup. In \cite{dmt2}, a gauge invariant formalism for target space entanglement in theories of matrices was developed in terms of suitable projection operators. The formalism of that paper cannot be applied to the present case in a straightforward fashion. However we hope to return to this problem in the near future.

\section{Discussion}

In this paper we have used the connection of Lowest Landau Level states to those of an auxiliary $1+1$ dimensional system to express the entanglement entropy of an arbitrary spatial subregion in $2+1$ dimensions in terms of the expectation value of the phase space density of this $1+1$ dimensional system. We showed that for the regime of interest where the subregion is much larger than the basic length scale of the theory and at large $N$ (the regime described in (\ref{regA})) the Thomas-Fermi approximation for the one dimensional phase space density is reliable.
This allows us to express the LLL state entanglement entropy in terms of integrals which we can perform analytically. We showed that in this leading order the result is a sum of a perimeter term with a coefficient which is independent of the shape of the subregion and a contribution from sharp corners. The latter is purely geometric (i.e. it does not involve any length scale) and depends only on the corner angle. Our results are in agreement with existing numerical results in the literature. There have been many works on entanglement entropy of integer quantum Hall states using the connection to noncommutative geometry and Chern-Simons theory, see e.g. \cite{nair} and references therein \footnote{See also discussions of entanglement entropies for fractional quantum Hall systems using Chern-Simons theory \cite{chern}}. It would be good to understand the connection of these approaches to ours, which directly deals with the fermionic field theory. 

When applied to the complex matrix model which follows from the 1/2-BPS sector of $\mathcal N=4$ SYM theory this entropy is the von Neumann entropy associated with the reduced density matrix which evaluates expectation values of a special class of operators. This is therefore akin to the kind of target space entanglement entropy discussed in \cite{dmt1}-\cite{hartnoll2}. Our discussion has been in a fixed gauge: we don't yet know what is the gauge-invariant description along the lines of \cite{dmt2}. However a gauge invariant definition should be possible. Discussions of the 2 matrix problem \cite{donos, cremonini} as well as using loop space ideas \cite{antalnew} could be useful in this regard.

A major outstanding question is to uncover the meaning of this quantity in the full ${\cal{N}}=4$ SYM theory and its gravitational bulk. As mentioned above, it is not clear if the results of the quadratic complex matrix model can be used to discuss a target space entanglement in the full theory, since this involves operators obtained from the 1/2-BPS operators by a projection. It is possible, however, that these results could give a qualitative guide. The question of relating to the supergravity bulk also deserves thought. 
While the SYM fields $X^I, I=1 \cdots 6$, represent the directions which are transverse to the stack of three branes which give rise to the $AdS_5 \times S^5$ geometry, the space of eigenvalues of these matrices may not be directly identified with this transverse space in the supergravity limit (See, however \cite{hanada}). In the present case, the question of identifying the bulk space with the space of eigenvalues of the matrix $Z$ is more complicated since $Z$ is not diagonalized by a gauge transformation, but instead reduced to a lower triangular matrix - this means that in this gauge neither $X^1$ nor $X^2$ are diagonal. Comparisons of the chiral primary correlators obtained from the matrix model with giant graviton correlators in supergravity \cite{pawel,skenderis} should be useful to understand this issue.
It is also important to understand the connection to other notions of entanglement (or entwinement) which involve internal spaces \cite{other1, other2, other3}. 
These issues will be addressed in a future communication.

Finally, we have dealt only with the ground state, whose dual is pure $AdS_5 \times S^5$. As we emphasized above, the use of phase space density is useful for excited states which are easily described in terms of filled regions of phase space, particularly when these regions are disconnected. 1/2-BPS states of this kind are gauge theory descriptions of giant gravitons and dual giants whose dual geometries are LLM geometries \cite{llm}. In fact the corresponding classical solutions are specified in terms of the classical phase space density of the one dimensional theory. This indicates that our procedure will be useful for such states. Extremal surfaces in such geometries have been constructed in \cite{balalawrence} and aspects of the entanglement structure of states of this type have been discussed in \cite{lin}. It will be interesting to relate these discussions with the entanglement in the fermionic theory.

\label{gaugegravity}

\section{Acknowledgements}

We would like to thank Pawel Caputa and Sandip Trivedi for discussions. We are grateful to Antal Jevicki for comments on an earlier draft which led to improvements, Satya Majumdar for correspondences clarifying several points and Gautam Mandal for discussions about many aspects of this paper. We thank William Witczak-Krempa and Benoit Estienne for comments about the first arXiv version. The work of S.R.D. is partially supported by U.S. National Science Foundation grants NSF/PHY-1818878 and NSF/PHY-2111673 and by the Jack and Linda Gill Chair Professorship. The work of S.H. is supported by the ERC Grant 787320 - QBH Structure. The work of S.L. is partially supported by NCN Sonata Bis 9 grants.

\section{Note added} While we were preparing this manuscript for submission to the arXiv, a related paper \cite{tsu2} appeared.

\newpage

\begin{center}
{\bf {\Large{Appendices}}}
\end{center}

\appendix

\section{One dimensional harmonic oscillator}
\label{apponedim}
In this appendix we provide the details of derivations of several equations in section \ref{2-1-2}.

We have the following entropy expression
\begin{equation}
\begin{split}\label{A-7}
S_{EE} &= {\pi^2\over3}\bigg({1\over 2\pi}\int_{-\infty}^{\infty}dp\int_A dxu(p,x,t) \cr
&-{1\over 4\pi^2} \int_{-\infty}^{\infty}dp_1\int_{-\infty}^{\infty}dp_2\int_A dx_1\int_Adx_2 e^{-i(p_2-p_1)(x_2-x_1)/\hbar}u\bigg(p_1,{x_1+x_2\over2},t\bigg) u\bigg(p_2,{x_1+x_2\over2},t\bigg)\bigg)\\
\end{split}
\end{equation}
This time we consider fermions in an excited state which corresponds to the geometry of an annulus of inner radius $r_1$ and outer radius $r_2=r_1+d$ in 1D phase space. This geometry is depicted in figure \ref{annulus}. As we can see there are two cases one can consider: 1) A subregion which is between $r_1$ and $r_2$ which is within the annulus and 2) a subregion which is in the region less than $r_1$. Our Wigner function is given by
\bea
u(p,x,t) &=& \Theta( r - r_1) \Theta( r_2 - r)
\eea
where 
\bea
r = \sqrt{p^2+\omega^2x^2}
\eea
Computing the area of the annulus yields the relation
\bea
r_2^2=r_1^2+2N\hbar\omega
\eea
and
\bea\label{d}
d = r_2-r_1 = \sqrt{r_1^2 + 2N\hbar\omega} - r_1
\eea

\subsection{Interval within region $r_1< \omega x < r_2$}
We first compute the entanglement entropy for a subregion within the annulus defined by $A$
\bea
A=\bigg\{ r_1/\omega + c -a\leq x\leq r_1/\omega + c + a\bigg\}
\eea 
where $c$ is the center of mass coordinate between $r_1$ and $r_2$ as measured from $r_1$ and $2\omega a$ is the size of the interval with both quantites measured along the $x$ axis. In order to keep the subregion within the annulus we require that
\bea\label{A-11}
a\leq c\leq d-a
\eea
Only when $r_1=0$ can we extend the range of $c$ down to $0$. 
We make the following coordinate changes
where a constant line cuts through the annulus at $\pm P_0(x)$
\bea\label{A-15}
P_{0}(x) &=& \sqrt{r_2^2 - \omega^2x^2} 
\eea

We consider the regime where the potential is slowly varying and therefore replace $P_0(x)\to P_0(r_1/\omega+c)$ for the first term in (\ref{A-7}) and $P_0(x_c)\to P_0(r_1/\omega+c)$ for the second term in (\ref{A-7}) where $x_c = {x_1+x_2\over2}$. This acts as a shift in variables. 
\bea
S_{EE} &=& {\pi^2\over3}\bigg({1\over 2\pi\hbar} \int_{ - a}^{ a} dx\int_{-P_0(r_1/\omega+c)}^{P_0(r_1/\omega+c)}dp \cr
&&-{1\over 4\pi^2\hbar^2}\int_{-a}^a dx_1\int_{-a}^{a}dx_2\int_{-P_0(r_1/\omega+c)}^{P_0(r_1/\omega+c)}dp_1\int_{-P_0(r_1/\omega+c)}^{P_0(r_1/\omega+c)}dp_2 e^{-i(p_2-p_1)(x_2-x_1)/\hbar}\bigg)
\cr
&=&{\pi\over3\hbar}2aP_0(r_1/\omega+c) - {1\over3} \int_{-a}^a dx_1\int_{-a}^{a}dx_2 {\sin^2((x_2-x_1)P_0(r_1/\omega+c)/\hbar)\over (x_2-x_1)^2}
\eea
Making the coordinate transformation
\bea\label{coord transf}
x_c = {x_1 + x_2\over2},\quad s = x_2 - x_1
\eea
we obtain
\bea
S_{EE}&=&{\pi\over3\hbar}2aP_0(r_1/\omega+c) - {4\over3} \int_0^{2a} ds\int_0^{a-{s\over2}}dx_c {\sin^2(sP_0(r_1/\omega+c)/\hbar)\over s^2}\cr
&=&{\pi\over3\hbar}2aP_0(r_1/\omega+c) + {1\over3}\big(1+ \gamma_E- \cos(4aP_0(r_1/\omega+c)/\hbar) - \text{Ci}(4aP_0(r_1/\omega+c)/\hbar)\cr
&&  +  \log(4aP_0(r_1/\omega+c)/\hbar)  - (4aP_0(r_1/\omega+c)/\hbar)\text{Si}(4aP_0(r_1/\omega+c)/\hbar)\big)
\eea
Expanding in the limit where $aP_0(r_1/\omega+c)/\hbar\gg1$ gives the leading order result
\bea
S_{EE} &=&{1\over3} ( 1 + \gamma_E + \log(4aP_0(r_1/\omega+c)/\hbar)) 
\eea

\subsection{Interval in region  $\omega x < r_1$}
Here we derive (\ref{2-30}) which is the entanglement entropy of a subregion within the region bounded by $r_1$. In this region constant $x$ lines cut the the boundary of the annulus in four places, $\pm P_0(x), \pm P_1(x)$. Where $P_0(x)$ is given by (\ref{A-15}) and $P_1(x)$ is given by (\ref{62-12}). Our subregion defined by 
\bea
A = x_0 - a \leq x \leq  x_0 + a
\eea
We again consider the fact that the potential is slowly varying. This allows to take $P_0(x)\to P_0(x_0)$ and $P_1(x)\to P_1(x_0)$ in the first term of (\ref{A-7}) and $P_0(x_c)\to P_0(x_0)$ and $P_1(x_c)\to P_1(x_0)$ in the second term of (\ref{A-7}). This gives
\begin{equation}
\begin{split}
S_{EE} &= {\pi^2\over3}\bigg[{1\over 2\pi\hbar} \int_{ - a}^{ a} dx\bigg(\int_{-P_0(x_0)}^{P_0(x_0)}dp - \int_{-P_1(x_0)}^{P_1(x_0)}dp\bigg)\cr
& -{1\over 4\pi^2\hbar^2} \int_{-a}^a dx_1\int_{-a}^{a}dx_2\bigg(\int_{-P_0(x_0)}^{P_0(x_0)}dp_1 - \int_{-P_1(x_0)}^{P_1(x_0)}dp_1\bigg)\bigg(\int_{-P_0(x_0)}^{P_0(x_0)}dp_2 - \int_{-P_1(x_0)}^{P_1(x_0)}dp_2\bigg) e^{-i(p_2-p_1)s/\hbar}\bigg]
\cr
\cr
&= {\pi\over3\hbar} 2a(P_0(x_0)-P_1(x_0))\cr
& - {1\over3}\int_{-a}^a dx_1\int_{-a}^{a}dx_2{(\sin((x_2-x_1)P_0(x_0) /\hbar) - \sin((x_2-x_1)P_1(x_1)/\hbar))^2\over (x_2-x_1)^2}
\end{split}
\end{equation}
Again changing coordinates we obtain
\bea
S_{EE}&=& {\pi\over3\hbar} 2a(P_0(x_0)-P_1(x_0))  - {4\over3} \int_0^{2a} ds\int_0^{a-{s\over2}}dx_c{(\sin(sP_0(x_0) /\hbar) - \sin(sP_1(x_0)/\hbar))^2\over s^2}\nonumber\\
\eea
Integrating the second term gives
\bea
S_{EE}&=& {\pi\over3\hbar} 2a(P_0(x_0)-P_1(x_0)) + {1\over3}\bigg(2 + 2 \gamma_E  -\cos(4aP_0(x_0)/\hbar)-\cos(4aP_1(x_0)/\hbar)\cr
&&-\text{Ci}(4aP_0(x_0)/\hbar) -\text{Ci}(4aP_1(x_0)/\hbar) + 2 \text{Ci}(2a(P_0(x_0) + P_1(x_0) )/\hbar)\cr
&& -  2 \text{Ci}(2a(P_0(x_0) -P_1(x_0) )/\hbar) + \log(4aP_0(x_0)/\hbar)  + \log(4aP_1(x_0)/\hbar)\cr
&& +  2\log((P_0(x_0) - P_1(x_0))/\hbar) -   2\log((P_0(x_0) + P_1(x_0))/\hbar)\cr
&& - 4\sin(2aP_0(x_0) /\hbar)\sin(2aP_1(x_0) /\hbar) - (4aP_0(x_0)\hbar)\text{Si}(4aP_0(x_0)/\hbar)\cr
&& - (4aP_1(x_0)\hbar)\text{Si}(4aP_1(x_0)/\hbar) + (4a(P_0(x_0)+P_1(x_0))/\hbar)\text{Si}(2a(P_0(x_0) + P_1(x_0))/\hbar)\cr
&& - (4a(P_0(x_0)-P_1(x_0))/\hbar)\text{Si}(2a(P_0(x_0) - P_1(x_0))/\hbar)\bigg)
\eea
Now we take the following limits, similar as before
\begin{equation}
\begin{split}
aP_0(x_0)/\hbar&\gg1\cr
aP_1(x_0)/\hbar&\gg1\cr
a(P_0(x_0)+P_1(x_0))/\hbar&\gg1\cr
a(P_0(x_0)-P_1(x_0))/\hbar&\gg1
\end{split}
\end{equation}
This gives the leading order expression recorded in (\ref{2-30}) 
\bea
S_{EE} = {1\over3}\bigg[2 + 2\gamma_E + \log\bigg({16a^2P_0(x_0)P_1(x_0)\over\hbar^2}\bigg({P_0(x_0) - P_1(x_0)\over P_0(x_0) + P_1(x_0)}\bigg)^2\bigg)\bigg]
\eea

\section{Evaluation of the kernel $\calk$}
\label{appkernel}

In this appendix we provide the derivation of (\ref{4-14}). Using (\ref{3-8}) and (\ref{3-8}) and (\ref{4-13}) in the expression for the kernel (\ref{4-8}) we get
\begin{equation}
\begin{split}
\mathcal{K} (x_1,x_2;v) =& \sum_{l=0}^\infty  \frac{1}{\ell \sqrt{l! \pi }} \left( \frac{x_1+i x_2}{\ell} \right)^l \exp \left(- \frac{x_1^2+x_2^2}{2 \ell^2}  \right) \times {\frac {1}{\sqrt {2^{l}\,l! \sqrt{\pi}}}} e^{-{\frac { v^{2}}{2 }}}\cdot H_{l}(v) \\
= & \pi^{-3/4} \ell^{-1} \exp \left(- \frac{x_1^2+x_2^2}{2 \ell^2} - {\frac { v^{2}}{2 }} \right) \sum_{l=0}^{\infty} \frac{1}{ l!}  \left( \frac{x_1+i x_2}{\sqrt{2} \ell} \right)^l H_l (v) \\
=& \pi^{-3/4} \ell^{-1} \exp \left(- \frac{x_1^2+x_2^2}{2 \ell^2} - {\frac { v^{2}}{2 }} \right) \exp \left\{ 2 v \left( \frac{x_1+i x_2}{\sqrt{2} \ell} \right) - \left( \frac{x_1+i x_2}{\sqrt{2} \ell} \right)^2 \right\} \\
= & \pi^{-3/4} \ell^{-1} \exp \left[- \frac{x_1^2+x_2^2}{2 \ell^2} - \left( \frac{x_1+i x_2}{\sqrt{2} \ell} \right)^2 \right] \exp \left[- {\frac { v^{2}}{2 }}+ 2 v \left( \frac{x_1+i x_2}{\sqrt{2} \ell} \right)  \right]
\end{split}
\label{k_cnc}
\end{equation}
which is (\ref{4-14}) in the text. Here we have used the generating fn. of the Hermite Poly.
\begin{equation}
e^{2xt-t^{2}}=\sum _{n=0}^{\infty }H_{n}(x){\frac {t^{n}}{n!}}
\end{equation}

\section{Asymptotic behaviour of the phase space density of 1d harmonic oscillator in large-$N$ limit}
In this section we determine the asymptotic form of the phase space density $\hat{u}(p,q,t)$ of 1D harmonic oscillator. Starting from (\ref{4-10}), (\ref{4-5}), and (\ref{4-13}), we find 
\begin{equation}
\hat{u} (p,x, 0)
=\frac{1}{\sqrt{\pi}} \int_{-\infty}^{+\infty} dy e^{i p y } e^{- \left(x^2+\frac{y^2}{4} \right)} \sum_{m,n=0}^{\infty} \frac{1}{\sqrt{2^m 2^n m! n!}} H_{m}\left(x+\frac{y}{2} \right) H_{n}\left(x-\frac{y}{2} \right) \hat{c}_{m,0}^{\dagger}\hat{c}_{n,0}
\end{equation}
By applying the identity
\begin{equation}
\int dx e^{-x^2} H_m (x+v) H_n (x+w) = 2^n \sqrt{\pi} m! w^{n-m} L_m^{n-m} (-2vw), ~~~m \le n
\label{tt}
\end{equation}
we can find the expectation value of $\hat{u}(p,q,t)$
\begin{equation}
u(p,x,0)  \equiv \langle F \left| \hat{u}(p,x,t) \right| F \rangle 
=2  e^{-\left(  x^2 +p^2 \right) } \sum_{n=0}^{N-1}(-1)^n   L_n \left( 2 ( x^2+ p^2) \right)
\label{uhol}
\end{equation}
where $L_n (x)$ is the Laguerre polynomial. We can learn from (\ref{uhol}) that in the $1D$ harmonic oscillator the expectation value of phase space density operator $\hat{u}$ is only a function of 
\begin{equation}
\rho \equiv r^2 \equiv x^2 +p^2
\label{defrho}
\end{equation}
A plot of $u(\rho)$ is given in figure \ref{uvn}.

\begin{figure}
\centering
\includegraphics[width=0.7\textwidth]{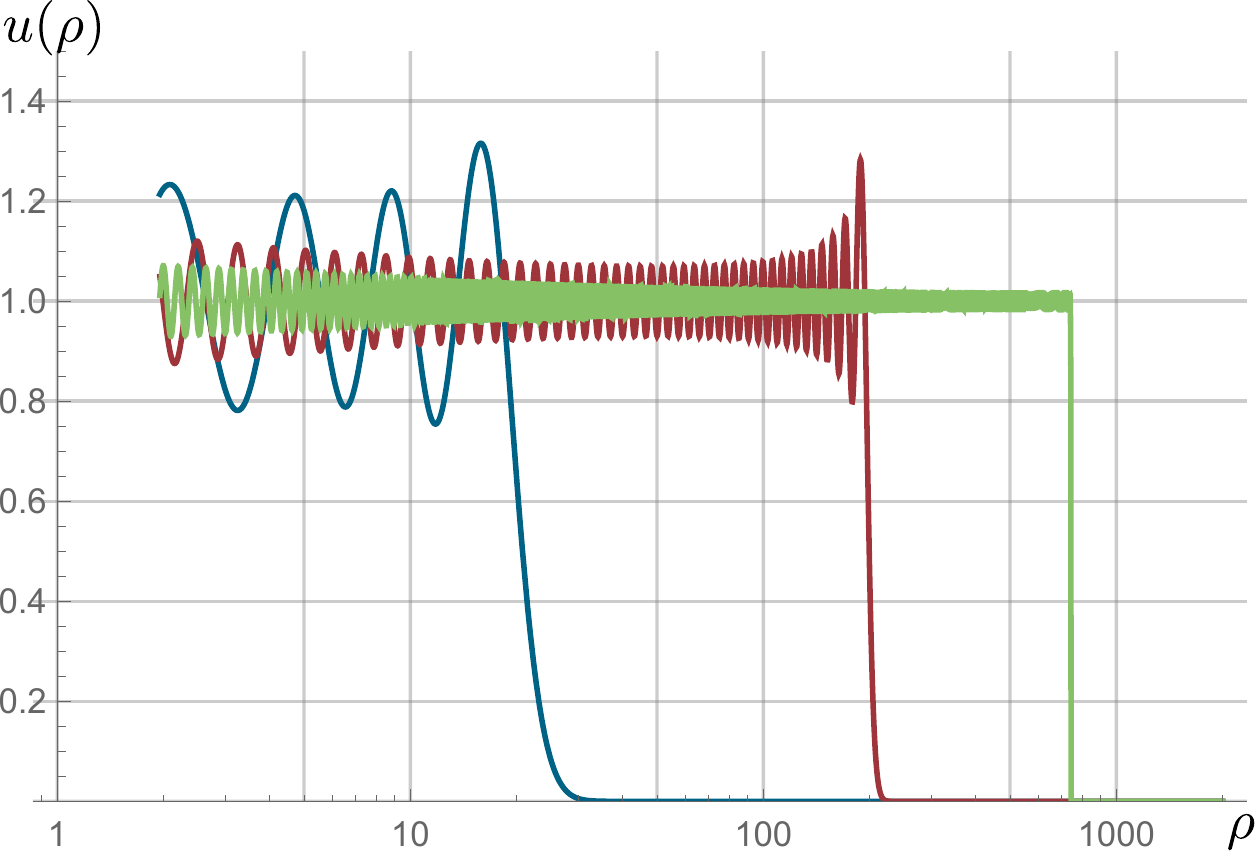}
\caption{The phase space density $u$ as a function of $\rho$ in (\ref{defrho}). The blue solid, red solid, and green solid lines show the behavior of $u$ in (\ref{uhol}) when $N=10, 100, 1000$, respectively.}
\label{uvn}
\end{figure}

\section{A direct calculation of correlation functions in large-$N$ limit} \label{appcrlfn}

In this section we evaluate the subleading contribution to correlation functions in (\ref{4-21}) in the large-$N$ limit via a direct calculation in two dimensions. From (\ref{3-8}) and (\ref{4-12}) we find
\bea\label{cf_dc}
&&\langle LLL | \psi^{\dagger}(x_1,x_2) \psi ( x_1', x_2') | LLL \rangle \cr
&&= \sum_{n,l =0}^{\infty} \phi_{n,0}^*(x_1,x_2) \phi_{l,0}(x_1',x_2') \langle LLL | \hat{c}_{n,0}^{\dagger} \hat{c}_{l,0} | LLL \rangle 
= \sum_{l=0}^{N-1} \phi_{n,0}^*(x_1,x_2) \phi_{l,0}(x_1',x_2') \cr
&&= \frac{1}{\pi \ell^2} \sum_{l=0}^{N-1} \frac{1}{l!}\left[ \left( \frac{x_1 -ix_2}{\ell} \right) \left( \frac{x_1'+i x_2'}{\ell} \right) \right]^l \exp\left( -\frac{x_1^2 + x_2^2 + x_1'^2 + x_2'^2}{2\ell^2} \right) \cr
&&= \frac{1}{\pi \ell^2} \sum_{l=0}^{\infty} \frac{1}{l!}\left[ \left( \frac{x_1 -ix_2}{\ell} \right) \left( \frac{x_1'+i x_2'}{\ell} \right) \right]^l \exp\left( -\frac{x_1^2 + x_2^2 + x_1'^2 + x_2'^2}{2\ell^2} \right) \cr
&&-\frac{1}{\pi \ell^2} \sum_{l=N}^{\infty} \frac{1}{l!}\left[ \left( \frac{x_1 -ix_2}{\ell} \right) \left( \frac{x_1'+i x_2'}{\ell} \right) \right]^l \exp\left( -\frac{x_1^2 + x_2^2 + x_1'^2 + x_2'^2}{2\ell^2} \right)
\eea

The leading term of (\ref{cf_dc}) is the Taylor series of an exponential. Thus
\begin{equation}
\begin{split}
& \frac{1}{\pi \ell^2} \sum_{l=0}^{\infty} \frac{1}{l!}\left[ \left( \frac{x_1 -ix_2}{\ell} \right) \left( \frac{x_1'+i x_2'}{\ell} \right) \right]^l \exp\left( -\frac{x_1^2 + x_2^2 + x_1'^2 + x_2'^2}{2\ell^2} \right) \\
& \qquad = \frac{1}{\pi \ell^2}  \exp\left( -\frac{(x_1-x_1')^2 + (x_2-x_2')^2}{2\ell^2} +i\frac{x_1x_2'-x_2x_1'}{\ell^2} \right) \\
\end{split}
\end{equation}
It is identical to the (\ref{4-21}). 
As for the second term of (\ref{cf_dc}), notice that
\begin{equation}
\begin{split}
\left| \sum_{l=N}^{\infty} \frac{1}{l!}\left[ \left( \frac{x_1 -ix_2}{\ell} \right) \left( \frac{x_1'+i x_2'}{\ell} \right) \right]^l  \right|  \qquad & \\
 \le   \exp \left[  \left|  \left( \frac{x_1-i x_2}{ \ell} \right)\left( \frac{ x_1'+i x_2'}{ \ell} \right) \right| \right] &  F \left(N;  \left|  \left( \frac{x_1-i x_2}{ \ell} \right)\left( \frac{ x_1'+i x_2'}{ \ell} \right) \right| \right) 
\end{split}
\end{equation}
where
\begin{equation}
\begin{split}
 & F \left(N;  \left|  \left( \frac{x_1-i x_2}{ \ell} \right)\left( \frac{ x_1'+i x_2'}{ \ell} \right) \right| \right)  \equiv \\
& \qquad \qquad \sum_{l=N}^{\infty} \frac{1}{l!} \left|  \left( \frac{x_1-i x_2}{ \ell} \right)\left( \frac{ x_1'+i x_2'}{ \ell} \right) \right|^l \exp \left[ - \left|  \left( \frac{x_1-i x_2}{ \ell} \right)\left( \frac{ x_1'+i x_2'}{ \ell} \right) \right| \right]
\end{split}
\end{equation}
is related to the cumulative distribution function of a Poisson distribution with parameter $\left|  \left( \frac{x_1-i x_2}{ \ell} \right)\left( \frac{ x_1'+i x_2'}{ \ell} \right) \right|$. Therefore a Chernoff bound argument shows that
\begin{equation}
\begin{split}
& F \left(N;  \left|  \left( \frac{x_1-i x_2}{ \ell} \right)\left( \frac{ x_1'+i x_2'}{ \ell} \right) \right| \right) \\
& \qquad \qquad  \leq {\frac {\left(e \left|  \left( \frac{x_1-i x_2}{ \ell} \right)\left( \frac{ x_1'+i x_2'}{ \ell} \right) \right| \right)^{N}}{N^{N}}} \exp \left[ - \left|  \left( \frac{x_1-i x_2}{ \ell} \right)\left( \frac{ x_1'+i x_2'}{ \ell} \right) \right| \right]
\end{split}
\label{pois_ineqn}
\end{equation}
for
\begin{equation}
\left|  \left( \frac{x_1-i x_2}{ \ell} \right)\left( \frac{ x_1'+i x_2'}{ \ell} \right) \right| < N
\end{equation}

Thus, if we consider the subregion $A$ much smaller than the complete region, we have the points $\vec{x} \equiv (x_1, x_2), \vec{x}' \equiv (x_1',x_2') \in A$ satisfy 
\begin{equation}
\left|\frac{\vec{x}}{\ell} \right|\left|\frac{\vec{x}'}{\ell}\right| = \left|  \left( \frac{x_1-i x_2}{ \ell} \right)\left( \frac{ x_1'+i x_2'}{ \ell} \right) \right| \ll N
\label{st_cd}
\end{equation}
Then by applying (\ref{pois_ineqn}), we find
\begin{equation}
\left| \sum_{l=N}^{\infty} \frac{1}{l!}\left[ \left( \frac{x_1 -ix_2}{\ell} \right) \left( \frac{x_1'+i x_2'}{\ell} \right) \right]^l  \right| \le   \frac{ e^{N} \left|  \left( \frac{x_1-i x_2}{ \ell} \right)\left( \frac{ x_1'+i x_2'}{ \ell} \right) \right|^{N} }{ N^{N}} \ll 1
\end{equation}
For example, if we take 
\begin{equation}
\left|\frac{\vec{x}}{\ell} \right|\left|\frac{\vec{x}'}{\ell}\right|  = \frac{N}{e^2} < N
\label{cf_cdt_e}
\end{equation}
we will have
\begin{equation}
\left| \sum_{l=N}^{\infty} \frac{1}{l!}\left[ \left( \frac{x_1 -ix_2}{\ell} \right) \left( \frac{x_1'+i x_2'}{\ell} \right) \right]^l  \right| \sim \mathcal{O}\left(  e^{-N} \right) \ll 1
\end{equation}
and therefore the second term of (\ref{cf_dc}) is estimated to be at the order 
\begin{equation}
\begin{split}
\frac{1}{\pi \ell^2} \sum_{l=N}^{\infty} \frac{1}{l!}\left[ \left( \frac{x_1 -ix_2}{\ell} \right) \left( \frac{x_1'+i x_2'}{\ell} \right) \right]^l \exp\left( -\frac{x_1^2 + x_2^2 + x_1'^2 + x_2'^2}{2\ell^2} \right) \sim \mathcal{O}\left( 2^{-N} e^{-N} \right)
\end{split}
\end{equation}
Indeed in most cases of (\ref{st_cd}), the LHS is even smaller than $\frac{N}{e^2}$, thus the second term can be neglected in the large $N$ limit. 

\section{Entanglement entropy for half-space}
\label{apphalf}

In this appendix we evaluate the integral in (\ref{55-1}).
The integration over $y_1,y_1^\prime$ is already finite in the $L_1 \rightarrow \infty$ limit, yielding
\ben
\int_0^\infty dy_1 \int_{-\infty}^0 dy_1^\prime 
{\rm exp} [ - (y_1 -y_1^\prime)^2] = \int_0^\infty dy_1 \int_0^\infty dy_1^\prime 
{\rm exp} [ - (y_1 + y_1^\prime)^2] = \frac{\sqrt{\pi}}{2}\int_0^\infty dy_1 [1-{\rm{erf}}(x)]=\frac{1}{2}
\label{55-22}
\een
In the remaining integrals, we perform 
perform the integral over $y_2^\prime$ first, which is finite in the $L_2 \rightarrow \infty$ limit. Finally we perform the integeral over 
\ben
\int_{-L_2}^{L_2} dy_2 \int_{-\infty}^\infty dy_2^\prime ~{\rm exp} [ - (y_2 -y_2^\prime)^2] = 2L_2 \sqrt{\pi}
\label{55-3}
\een
Using (\ref{55-22}) and (\ref{55-3}) in (\ref{55-1}) we get the final result (\ref{55-4}).

\section{A generic smooth entangling curve}
\label{general}

In this section we determine under which condition we can apply (\ref{5-14}) to a general smooth entangling curve $\partial A$. We know that for an arbitrary point on $\partial A$, which we denote by $P=(x_1,x_2)$, there exists an inscribed circle with radius $R_P = K_P^{-1}$, where $K_P$ is the signed curvature at $P$. In particular, this inscribed circle can be described by an implicit function parametrized by the position of $P$, $\vec{r}_P$, outward normal vector to $\partial A$ at $P$, $\hat{n}_P$, and radius (inverse curvature at $P$) $R_P$:
\begin{equation}
\left| \vec{r} - \left( \vec{r}_P - R_P \hat{n}_P \right) \right| = R_P \Leftrightarrow
\left( \vec{r} -\vec{r}_P \right)^2 + 2 R_P \hat{n}_P \cdot \left( \vec{r} -\vec{r}_P \right) =0
\end{equation}
We can choose this inscribed circle as a good approximation of real curve $\partial A$ around $P$. Then similar to (\ref{5-10}), we have
\begin{equation}
\vec{n}_x \cdot \nabla_x \mathbf{1}_{\vec{x} \in A^c_R} \big|_P = \delta \left( r - R_P \right) 
\end{equation}
where 
\begin{equation}
r = \left| \vec{r} - \left( \vec{r}_P - R_P \hat{n}_P \right) \right|  
\end{equation}
The same argument can be made for another point in the neighborhood of $P$ at $\partial A$ denoted by $P'=(x_1^P + \delta x_1 , x_2^P + \delta x_2)$ (see figure \ref{fig_cnn}), and we obtain $\vec{n}_x \cdot \nabla_x \mathbf{1}_{\vec{x} \in A^c_R} \big|_{P'}$. 
Then we can expand $\vec{n}_x \cdot \nabla_x \mathbf{1}_{\vec{x} \in A^c_R} \big|_{P'}$ around $\vec{n}_x \cdot \nabla_x \mathbf{1}_{\vec{x} \in A^c_R} \big|_P$ and obtain
\begin{equation}
\begin{split}
\delta \left( r' - R_{P'} \right) 
=& \delta \left( r - R_P \right) + \left[ (r'- r) - (R_{P'}-R_P) \right] \delta'  \left( r - R_P \right) \\
&+ \frac{1}{2} \left[ (r'- r) - (R_{P'}-R_P) \right]^2 \delta''  \left( r - R_P \right)
\end{split}
\label{epddel}
\end{equation}

\begin{figure}
\centering
\includegraphics[width=0.45\textwidth]{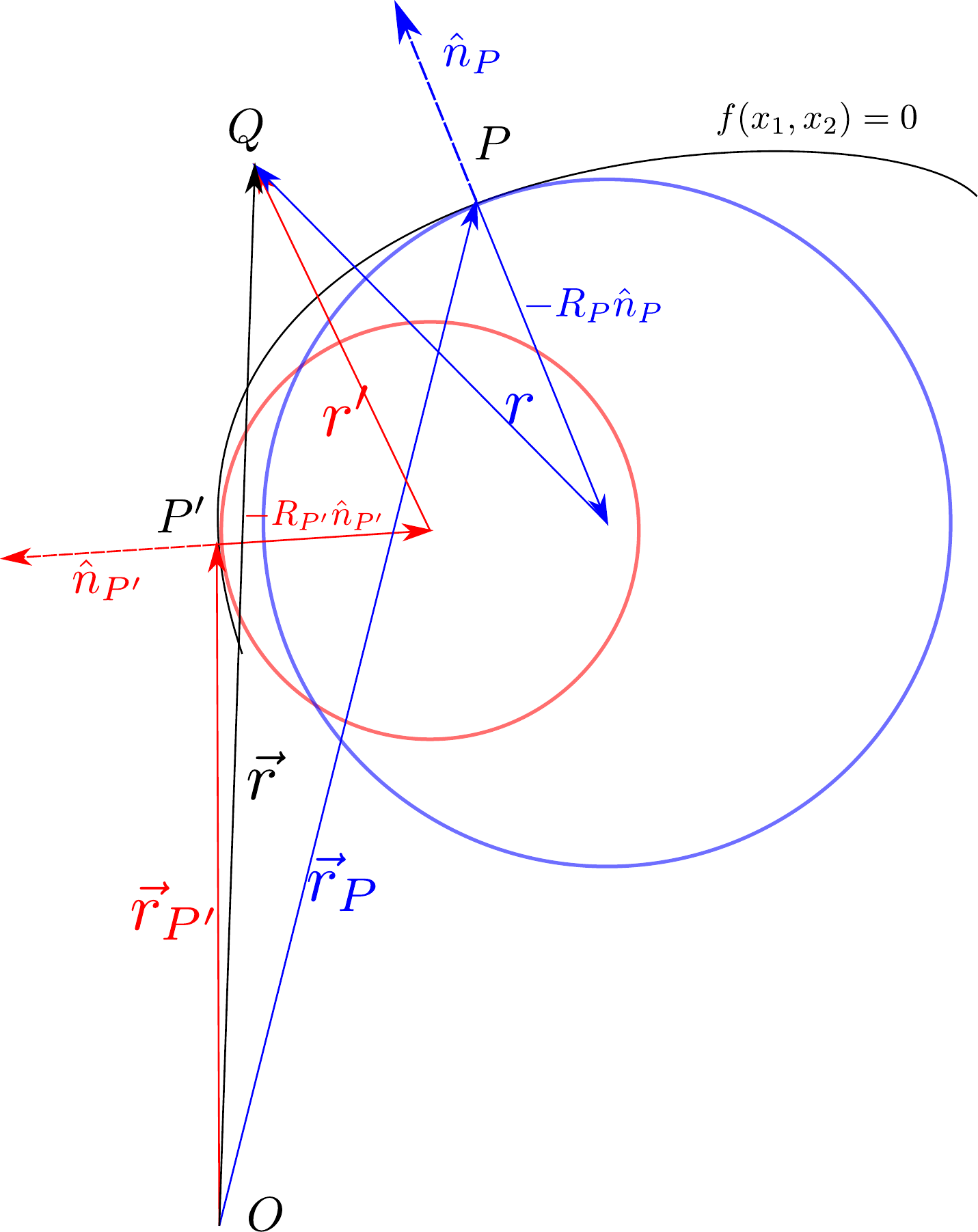}
\caption{Relations among inscribed circles, vectors, and etc.}
\label{fig_cnn}
\end{figure}
Below we evaluate $(r'- r) - (R_{P'}-R_P)$. Without loss of generality, we consider 
\begin{equation}
\partial A: f(x_1,x_2) = x_2 -g(x_1) = 0, \qquad g'(x_1)=0
\end{equation}
and therefore the radius takes a simple form 
\begin{equation}
R_P =- \frac{1}{g''}
\end{equation}
Thus, via the intermediate step
\begin{equation}
\begin{split}
& \left| \vec{r}_{P'} - \left( \vec{r}_P - R_P \hat{n}_P \right) \right|^2 
= \left( \vec{r}_{P'} - \vec{r}_P \right)^2  + 2 \left( \vec{r}_{P'} - \vec{r}_P \right) \cdot  R_{P} \hat{n}_{P} + R_P^2 \\
=&  \delta x_1^2 + \left(\frac{1}{2} g''(x_1) \delta x_1^2  +\mathcal{O}(\delta x_1^3)\right)^2 + 2 \frac{1+g'^2}{-g''} \left[ -g' \times \delta x_1 + 1 \times \left( g'\delta x_1 + \frac{1}{2} g'' \delta x_1^2 +\frac{1}{6} g''' \delta x_1^3 \right) \right]  + R_P^2 \\
=& R_P^2 - \frac{1}{3} \frac{g'''}{g''} \delta x_1^3 + \left( \frac{1}{4} g''^2 - \frac{1}{12} \frac{g^{(4)}}{g'' } \right) \delta x_1^4 + \mathcal{O}(\delta x_1^5) \\
=& R_P^2 + \frac{1}{3} \partial_1 \log R_P \delta x_1^3 -  \frac{1}{12} \left[ (\partial_1 \log R_P)^2 - \partial_1^2 \log R_P \right]  \delta x_1^4 + \mathcal{O}(\delta x_1^5) \\
\end{split}
\label{rpprime}
\end{equation}
where $g^{(p)}$ denotes the $p$'th derivative w.r.t $r$, we find
\begin{equation}
\begin{split}
|r'-r| =& \left| \left| \vec{r} - \left( \vec{r}_{P'} - R_{P'} \hat{n}_{P‘} \right) \right| - \left| \vec{r} - \left( \vec{r}_P - R_P \hat{n}_P \right) \right|  \right| \\
\le & \left| \left( \vec{r}_{P'} - R_{P'} \hat{n}_{P‘} \right) - \left( \vec{r}_P - R_P \hat{n}_P \right) \right| \\
=& \left[ \left( \vec{r}_{P'} - \vec{r}_P \right)^2  -2 \left( \vec{r}_{P'} - \vec{r}_P \right) \cdot \left(  R_{P'} \hat{n}_{P'} - R_{P} \hat{n}_{P} \right)
+\left(  R_{P'} \hat{n}_{P'} - R_{P} \hat{n}_{P} \right)^2  \right]^{1/2} \\
=& \left[ -\left( \vec{r}_{P'} - \vec{r}_P \right)^2  
+\left(  R_{P'} \hat{n}_{P'} - R_{P} \hat{n}_{P} \right)^2 +\mathcal{O}(\delta x_1^3) \right]^{1/2}
\end{split}
\label{triine}
\end{equation}
where we have applied the triangle inequality and the symmetry between $P \leftrightarrow P'$.

Furthermore, a similar calculation to (\ref{rpprime}) yields
\begin{equation}
\begin{split}
\hat{n}_P \cdot \hat{n}_{P'} =&
1- \frac{1}{2} k_P^2 (1+g'^2) \delta x_1^2 +\mathcal{O}(\delta x_1^3) \\
=& 1- \frac{1}{2} k_P^2 \left( \vec{r}_{P'} - \vec{r}_P \right)^2 \\
& + \frac{1}{2} \frac{\partial_1 R_P}{R_P^3}\delta x_1^3 - \left[ \frac{1}{8} \frac{1}{R_P^4} + \frac{11}{24} \frac{(\partial_1 R_P)^2}{R_P^4}-\frac{1}{6} \frac{\partial_1^2 R_P}{R_P^3}\right]  \delta x_1^4 +\mathcal{O}(\delta x_1^5) 
\end{split}
\label{npnpp}
\end{equation}
Then because
\begin{equation}
\begin{split}
\left(  R_{P'} \hat{n}_{P'} - R_{P} \hat{n}_{P} \right)^2 = \left(  R_{P'}  - R_{P}  \right)^2 + 2R_P R_{P'} \left(1 - \hat{n}_P \cdot \hat{n}_{P'} \right)
\end{split}
\label{nna0}
\end{equation}
we have
\begin{equation}
\begin{split}
& \left| \left( \vec{r}_{P'} - R_{P'} \hat{n}_{P'} \right) - \left( \vec{r}_P - R_P \hat{n}_P \right) \right| \\
=& \left[ -\left( \vec{r}_{P'} - \vec{r}_P \right)^2  +  \left(  R_{P'}  - R_{P}  \right)^2 + R_P R_{P'} \times k_P k_{P'} \left( \vec{r}_{P'} - \vec{r}_P \right)^2 + \mathcal{O}(\delta x_1^3) \right]^{1/2} \\
=& \left|  R_{P'}  - R_{P}  \right| +  \left|  R_{P'}  - R_{P}  \right|^{-1} \mathcal{O}(\delta x_1^3)
\end{split}
\label{ineqrr3}
\end{equation}
where
\begin{equation}
\mathcal{O} (\delta x_1^3) \sim \frac{\partial_1 R_P}{R_P} \delta x_1^3 + \mathcal{O} (\delta x_1^4)
\end{equation}
according to (\ref{rpprime}) and (\ref{npnpp}).

Notice that 
\begin{equation}
\left|  R_{P'}  - R_{P}  \right| \sim \partial_1 R_P \delta x_1 + \frac{1}{2}\partial_1^2 R_P \delta x_1^2 +\mathcal{O}(\delta x_1^3)
\end{equation}
Moreover, we can express $\delta x_1$ in terms of $\delta \phi$, which is the angular variation from $P$ to $P'$ if we choose the center of inscribed circle at $P$ as the origin of a polar coordinate system. In particular,
\begin{equation}
\delta x_1 \le   R_P \delta \phi + \left| \vec{r}_{P'} - \left( \vec{r}_P - R_P \hat{n}_P \right) \right| -R_P = R_P \delta \phi + \mathcal{O} \left({\partial_{\phi}  R_P} \delta \phi^3 \right)
\end{equation}
Thus, we eventually obtain
\begin{equation}
\begin{split}
( r'- r) - (R_{P'}-R_{P}) 
& \le \left| \left( \vec{r}_{P'} - R_{P'} \hat{n}_{P‘} \right) - \left( \vec{r}_P - R_P \hat{n}_P \right) \right| + \left| R_{P'}-R_{P} \right| \\
& = 2\left|  R_{P'}  - R_{P}  \right| + \left|  R_{P'}  - R_{P}  \right|^{-1} \mathcal{O}(\delta x_1^3) \\
& \sim 2\partial_{\phi} R_P \delta \phi +\mathcal{O} \left( {R_P} \delta \phi^2 , \partial_{\phi}^2 R_P  \delta \phi^2 \right) 
\end{split}
\label{cfest}
\end{equation}

Because the Laplace operator turns into
\begin{equation}
\nabla^2 \big|_P = \partial_{r}^2 + \frac{1}{R_P} \partial_r + \frac{1}{R_P^2} \partial_{\phi}^2
\end{equation}
at $P$, we plug (\ref{cfest}) back into (\ref{epddel}) and find that
\begin{equation}
\begin{split}
& \nabla^2 \left[ \vec{n}_x \cdot \nabla_x \mathbf{1}_{\vec{x} \in A^c_R} \right] \bigg|_P \\ =&~ \partial_r^2 \delta \left( r - R_P \right) \\
& + \mathcal{O} \left(\left(\frac{\partial_{\phi} R_P}{R_P} \right)^2 \right) \partial_r^2 \delta \left( r - R_P \right) + \mathcal{O} \left(\frac{1}{R_P} \right) \partial_r \delta \left( r - R_P \right) + \mathcal{O} \left(\frac{\partial_{\phi}^2 R_P}{R_P^2}\right) \partial_r \delta \left( r - R_P \right) 
\end{split}
\end{equation}
This implies that (\ref{5-14}) is true for a general smooth entangling curve $\partial A$ when
\begin{equation}
\operatorname{max} \left\{ \left| \partial_{\phi} R \right|, 1 \right\} \ll R
\label{pl_cdt}
\end{equation}

In terms of curvature and the symbols in section \ref{smooth}, (\ref{pl_cdt}) is equivalent to
\begin{equation}
\left| \partial_{\sigma} K (\sigma) \right| \ll \left| K (\sigma) \right| \ll 1  
\label{pl_cdtk}
\end{equation}
We need to point out here that the conclusion is true not only for $\nabla^2 \left[ \vec{n}_x \cdot \nabla_x \mathbf{1}_{\vec{x} \in A^c_R} \right]$, but also for $\nabla^{2l} \left[ \vec{n}_x \cdot \nabla_x \mathbf{1}_{\vec{x} \in A^c_R} \right]$. The argument somewhat follows the idea of mathematical induction: 
Above we have figured out the case when $l=1$. 
Now if for $l=k$ we have
\begin{equation}
 \nabla^{2l} \left[ \vec{n}_x \cdot \nabla_x \mathbf{1}_{\vec{x} \in A^c_R} \right]  \approx \partial_r^{2l} \delta \left( r - K^{-1} (\sigma) \right) 
\label{pl2l}
\end{equation}
then for $l=k+1$, we can adjust the argument by replacing the delta functions in (\ref{epddel}) by $\partial_r^{2k} \delta ( r -R_P)$ or $\partial_{r'}^{2k} \delta ( r' -R_{P'})$, respectively. Since $(r'- r) - (R_{P'}-R_P)$ is the same as for $l=1$, nothing else needs to be changed. Thus we see that (\ref{pl2l}) is true when $l=k+1$ under the condition (\ref{pl_cdtk}). 

In conclusion, (\ref{pl_cdtk}) is the condition for the validity of (\ref{5-14}) in the case of a general smooth entangling curve.


\section{Entanglement entropy of triangle-like subregion}
\label{apptriangle}

In this section we derive expression (\ref{6-9}) which gives the entanglement entropy of a subregion in the shape of an isosceles triangle where the height $h \gg 1$ in units of $\ell$. 
An isosceles triangular subregion can parameterized in the following way
\bea
A = \bigg\{ (x,y)\bigg| -kx\leq y \leq k x, 0 \leq x \leq h \bigg\}
\eea
To compute the entanglement entropy of such a region we use the expression (\ref{5-3}) which contains two terms. So we have
\bea
S_A=S_{1} + S_{2}
\eea
where
\bea
S_{1}&=&{\pi^2\over3}\operatorname{tr}C_A \cr
S_{2}&=&-{\pi^2\over3}\operatorname{tr}C^2_A
\eea
\subsection*{Computing $S_{1}$}
 Here we compute the leading order term, $S_1$, which is proportional to the number density of fermions. 
 Using the expression (\ref{4-21}) we obtain the following for $S_1$
\bea\label{S1}
{3\over\pi^2}S_1 &=& \operatorname{tr}C_A\cr
&=& \frac{1}{\pi l^2_B} \int_A dxdy  
= \frac{1}{\pi l^2_B} \int_0^h dx \int_{-k x}^{k x} dy 
   = \frac{h^2k}{\pi \ell^2}
\eea
\subsection*{Computing $S_2$}
For the second term, $S_2$, we integrate $y_1, y_2$ first and obtain
\begin{equation}
\begin{split}
{3\over\pi^2}S_2 =&-\operatorname{tr}C^2_A\\
=& -\frac{1}{\pi^2 \ell^4} \int_0^h d x_1 \int_{-k x_1}^{k x_1} d y_1\int_0^h d x_2 \int_{-k x_2}^{k x_2} d y_2 ~ e^{-\frac{(x_1-x_2)^2}{\ell^2}} e^{-\frac{(y_1-y_2)^2}{\ell^2}} \\
= & -\frac{1}{\pi^2 } \int_0^{h\ell^{-1}} d x_1 d x_2 ~  e^{-(x_1-x_2)^2} \\
& \times \bigg\{ -e^{-k^2 (x_1-x_2)^2}  +  e^{- k^2(x_1+ x_2)^2} \\
& \qquad +\sqrt{{\pi}}\left[ - k (x_1 -x_2) \operatorname{erf} {k (x_1-x_2)}    +  k (x_1 +x_2) \operatorname{erf} {k(x_1+x_2)}\right]
\bigg\} \\
\end{split}
\end{equation}
Here in the last equation we have rescaled the $x$'s and $y$'s by $\ell$ for convenience.
Making the change of variables
\begin{equation}
x_{\pm} = x_1 \pm x_2
\end{equation}
gives
\bea
{3\over\pi^2}S_2
&= & -\frac{1}{2\pi^2} \left(  \int_0^{h\ell^{-1}} d x_- \int_{x_-}^{2{h\ell^{-1}}-x_-} d x_+ + \int_{-{h\ell^{-1}}}^{0} d x_- \int_{-x_-}^{2{h\ell^{-1}}+x_-} d x_+  \right)~  e^{-x_-^2} \cr
&& \times \left\{ -e^{-k^2 x_-^2}  +  e^{- k^2 x_+^2} 
+\sqrt{\pi}\left[ - k x_- \operatorname{erf} k x_-    +  k x_+ \operatorname{erf} kx_+\right]  \right\} \cr
&=& -\frac{1}{\pi^2}    \int_0^{h\ell^{-1}} d x_- \int_{x_-}^{2{h\ell^{-1}}-x_-} d x_+ ~  e^{-x_-^2} \cr
&& \times \left\{ -e^{-k^2 x_-^2}  +  e^{- k^2 x_+^2} 
+\sqrt{\pi}\left[ - k x_- \operatorname{erf} k x_-    +  k x_+ \operatorname{erf} kx_+\right]  \right\} 
\eea
Integrating $x_+$ first yields
\begin{equation}
\begin{split}
 {3\over\pi^2}S_2 
=&  -\frac{1}{\pi^2}  \int_0^{h\ell^{-1}} d x_- ~  e^{-x_-^2} \\
& \times \left\{ - 2({h\ell^{-1}}-x_-) e^{-k^2 x_-^2}+ \frac{1}{2} (2{h\ell^{-1}}-x_-) e^{-k^2(2{h\ell^{-1}}-x_-)^2} - \frac{1}{2} x_-e^{-k^2 x_-^2}      \right. \\
&\left.  -2\sqrt{\pi}  k x_- (h\ell^{-1}-x_-) +\frac{1}{4k}\left[ \sqrt{\pi}\left( 1+2k^2(2h\ell^{-1}-x_-)^2 \right)  - \sqrt{\pi}\left( 1+2k^2 x_-^2 \right)  \right] \right.\\
& \left. +2\sqrt{\pi}  k x_- (h\ell^{-1}-x_-) \operatorname{erfc} k x_-  - \frac{1}{4k}\bigg[ \sqrt{\pi}\left( 1+2k^2(2h\ell^{-1}-x_-)^2 \right) \operatorname{erfc} k(2h\ell^{-1}-q_-) \right.\cr
&\left. - \sqrt{\pi}\left( 1+2k^2 x_-^2 \right) \operatorname{erfc} k x_- \bigg]
\right\} \\
\equiv \mathfrak{I} +\mathfrak{K}
\end{split}
\end{equation}
where
\bea
\mathfrak{I}&\equiv& -\frac{1}{\pi^2}  \int_0^{h\ell^{-1}} d x_- ~  e^{-x_-^2}  \cr
&& \times \left\{ - 2({h\ell^{-1}}-x_-) e^{-k^2 x_-^2}+ \frac{1}{2} (2{h\ell^{-1}}-x_-) e^{-k^2(2{h\ell^{-1}}-x_-)^2} - \frac{1}{2} x_-e^{-k^2 x_-^2}      \right. \cr
&& \left.  -2\sqrt{\pi}  k x_- (h\ell^{-1}-x_-) +\frac{1}{4k}\left[ \sqrt{\pi}\left( 1+2k^2(2h\ell^{-1}-x_-)^2 \right)  - \sqrt{\pi}\left( 1+2k^2 x_-^2 \right)  \right] \right\} \\
\mathfrak{K}&\equiv&  -\frac{1}{\pi^2}  \int_0^{h\ell^{-1}} d x_- ~  e^{-x_-^2}  \cr
&& \times \left\{ 2\sqrt{\pi}  k x_- (h\ell^{-1}-x_-) \operatorname{erfc} k x_-  - \frac{1}{4k}\bigg[ \sqrt{\pi}\left( 1+2k^2(2h\ell^{-1}-x_-)^2 \right) \operatorname{erfc} k(2h\ell^{-1}-q_-) \right.\cr
&&\left. - \sqrt{\pi}\left( 1+2k^2 x_-^2 \right) \operatorname{erfc} k x_- \bigg]
\right\} 
\eea
For convenience let $h\ell^{-1} \to h$ below. we will restore $\ell$ eventually.
The term $\mathfrak{I}$ can be integrated straightforwardly. We get
\begin{equation}
\begin{split}
\mathfrak{I} 
=& -\frac{1}{\pi^2}\left\{ \frac{1}{2}(1+2h^2)k \pi \operatorname{erf} h + \frac{1}{4(1+k^2)}\left[ 3-e^{-4k^2h^2}-2 e^{-(1+k^2)h^2} \right] 
-2\sqrt{\pi}hk\right.\\
&\left. +  \sqrt{\pi} e^{-h^2}hk  + \frac{1}{2}\sqrt{ \pi} (1+k^2)^{-3/2} \left[  e^{-\frac{4h^2k^2}{1+k^2}} h \operatorname{erf} \frac{2 hk^2}{\sqrt{1+k^2}} - 2 h \operatorname{erf} h\sqrt{1+k^2} \right.\right.\\
&\left.\left. - 2 h k^2 \operatorname{erf} h\sqrt{1+k^2} + e^{-\frac{4h^2k^2}{1+k^2}}h \operatorname{erf} \frac{h-hk^2}{\sqrt{1+k^2}} \right] \right\} \\
=
& -\frac{1}{\pi^2} \left\{ \frac{1}{2}(1+2h^2)k \pi + \frac{3}{4(1+k^2)} -2\sqrt{\pi} hk -\sqrt{\pi} (1+k^2)^{-1/2}h \right\}\cr
& + \mathcal{O}(e^{-h^2}) \\
\end{split}
\label{I}
\end{equation}
in the limit where $h \gg 1$.
Now consider the term $\mathfrak{K}$. To evaluate $\mathfrak{K}$ it will be helpful to recall some useful relations.
\begin{equation}
\int_0^{\infty} e^{-(at^2+2bt+c)} dt = \frac{1}{2}\sqrt{\frac{\pi}{a}} e^{\frac{b^2-ac}{a}} \operatorname{erfc} \frac{b}{\sqrt{a}}, ~~~\Re a >0
\end{equation}
Taking
\begin{equation}
a=1,~~~c=0,~~~b=kx_- {\text{ or }} b= k(2h-x_-)
\end{equation}
gives
\begin{eqnarray}
\int_0^{\infty} e^{-t^2-2 kx_- t}dt &=& \frac{1}{2} \sqrt{\pi} e^{k^2 x_-^2} \operatorname{erfc} kx_- \\
\int_0^{\infty} e^{-t^2-2 k(2h-x_-) t}dt &=& \frac{1}{2} \sqrt{\pi} e^{k^2 (2h-x_-)^2} \operatorname{erfc} k(2h-x_-)
\end{eqnarray}
Now applying these relations to $\mathfrak{K}$ we obtain
\begin{equation}
\begin{split}
\mathfrak{K} =& -\frac{1}{\pi^2}  \int_0^{h} d x_- ~  e^{-x_-^2}  \\
& \times \left\{ 4 k x_- (h-x_-) e^{-k^2x_-^2}\int_0^{\infty} e^{-t^2-2 kx_- t}dt  \right. \\
 & \qquad \left.- \frac{1}{2k}\bigg[ \left( 1+2k^2(2h-x_-)^2 \right) e^{-k^2 (2h-x_-)^2}  \int_0^{\infty} e^{-t^2-2 k(2h-x_-) t}dt \right.\cr
 & \qquad \qquad \left. - \left( 1+2k^2 x_-^2 \right)e^{-k^2x_-^2} \int_0^{\infty} e^{-t^2-2 kx_- t}dt  \bigg]
\right\} \\
=& -\frac{1}{\pi^2}\int_0^{\infty}  dt  ~ e^{-t^2} \times \int_0^h d x_-~  e^{-x_-^2} \\
& \times \left\{ 4  k x_- (h-x_-) e^{-k^2 x_-^2}  e^{-2 kx_- t}+ \frac{1}{2k}\left( 1+2k^2 x_-^2 \right) e^{-k^2 x_-^2}  e^{-2 k x_- t} \right\} + \mathcal{O}(e^{-k^2h^2}) \\
\end{split}
\end{equation}
Given that the second term is $\sim \mathcal{O}(e^{-k^2h^2})$, we have ignored it in the last line. Now integrate $x_-$ and again simplify our result approximating higher terms, i.e. $\operatorname{erf}\frac{hk^2+h+kt}{\sqrt{1+k^2}} \to 1, e^{-h(hk^2+h+2kt)} \to 0$. Firstly integrating over $x_-$ yields
\bea\label{K}
\mathfrak{K} 
&=& -\frac{1}{\pi^2}\int_0^{\infty}  dt  ~ e^{-t^2} \cr
&& \times \left\{ \frac{1}{4k} ( 1+k^2)^{-5/2} \sqrt{\pi} \left[ -6 k^4 t^2 -8h(1+k^2)k^3 t -3k^2(1+k^2)+(1+k^2)^2 \right]\right.\cr
 &&~~~~~\left.e^{\frac{k^2 t^2}{1+k^2}}\operatorname{erfc} \frac{ k t}{\sqrt{1+k^2}}  \right. \cr
&&~~~~~ \left. +\frac{1}{4(1+k^2)^{2}} \left[ 8hk(1+k^2) +6k^2 t \right]  \right\} 
\eea
Integrating (\ref{K}) over the final variable $t$ gives
\begin{equation}
\begin{split}
\mathfrak{K}
=& -\frac{1}{\pi^2} \left\{ - \sqrt{\pi}\frac{hk^2}{\sqrt{1+k^2}}  + \sqrt{\pi} k h  + \frac{3k^2}{4(1+k^2)} + \frac{1}{4}  \left( \frac{1}{k}-3k \right) \cot^{-1} k \right\}
\end{split}
\label{K final}
\end{equation}
Combining the results for $\mathfrak{I}$ and $\mathfrak{K}$ in  (\ref{I}) and (\ref{K final}) respectively yields
\bea\label{S2}
{3\over\pi^2}S_2 &=&~ \mathfrak{I} + \mathfrak{K} \\ 
&=& -\frac{1}{\pi^2} \left\{ \frac{1}{2}(1+2h^2)k \pi + \frac{3}{4(1+k^2)} -2\sqrt{\pi} hk -\sqrt{\pi} (1+k^2)^{-1/2}h \right\} \cr
&&  -\frac{1}{\pi^2} \left\{ - \sqrt{\pi}\frac{hk^2}{\sqrt{1+k^2}}  + \sqrt{\pi} k h  + \frac{3k^2}{4(1+k^2)} + \frac{1}{4}  \left( \frac{1}{k}-3k \right) \cot^{-1} k \right\} \cr
&&+ \mathcal{O}(e^{-h^2},e^{-k^2 h^2}) \cr
&=& -\frac{1}{\pi} kh^2\ell^{-2} + \pi^{-3/2}\ell^{-1} \left[ hk+h\sqrt{1+k^2} \right] -\left[ \frac{k}{2\pi} + \frac{3}{4\pi^2} + \frac{1}{4\pi^2} \left( \frac{1}{k}-3k \right) \cot^{-1} k \right]\nonumber\\
\eea
where in the last line we have restored $\ell$.
~\\
Combining expressions for $S_1$ and $S_2$ given in (\ref{S1}) and (\ref{S2}) respectively yields an entropy of 
\begin{equation}
S_A ={\pi^2\over3}\bigg\{ (2\pi)^{-3/2} \left(\frac{\ell}{\sqrt{2}} \right)^{-1} \cdot \left[ 2hk+2h\sqrt{1+k^2} \right] -\left[ \frac{k}{2\pi} + \frac{3}{4\pi^2} + \frac{1}{4\pi^2} \left( \frac{1}{k}-3k \right) \cot^{-1} k \right]\bigg\}
\end{equation}
Furthermore, since $k$ is the slope of the edge, we can write it as
\begin{equation}
k= \tan \frac{\alpha}{2}
\end{equation}
where $\alpha$ is the vertex angle. As a result the corner term, which we define as $\gamma$ becomes
\bea
\gamma &\equiv& -{\pi^2\over3}\left[ \frac{1}{2\pi}\tan \frac{\alpha}{2} + \frac{3}{4\pi^2} + \frac{1}{4\pi^2} \left( \cot \frac{\alpha}{2}-3 \tan \frac{\alpha}{2} \right) \left( \frac{\pi}{2} - \frac{\alpha}{2} \right) \right]\cr
&=& -\frac{1}{12 \sin \alpha} \left[ \pi +\alpha +3\sin \alpha - 2\alpha \cos \alpha \right]
\label{gamma}
\eea
The entropy becomes
\bea
S_A &=&{\pi^2\over3}\bigg\{ (2\pi)^{-3/2}\left(\frac{\ell}{\sqrt{2}} \right)^{-1}\cdot \left[ 2hk+2h\sqrt{1+k^2} \right] -\frac{1}{4\pi^2 \sin \alpha} \left[ \pi + \alpha +3\sin \alpha - 2\alpha \cos \alpha \right]\bigg\}\cr
&=& \frac{\sqrt{\pi}}{6}\calp (k;h) -\frac{1}{12 \sin\alpha} \left[ \pi + \alpha +3\sin \alpha - 2\alpha \cos \alpha \right]
\eea
where $\calp (k;h) = \mathcal{P}(\alpha;h)$ is the perimeter term
\bea
\calp (k;h) &=& 2h(k+\sqrt{1+k^2})\cr
&=& 2h\bigg(\tan \frac{\alpha}{2}+\sec{\alpha\over2}\bigg) =\mathcal{P}(\alpha;h)
\eea

\section{Entanglement entropy of pie-slice subregion}
\label{apppie}

In this section we study the entanglement entropy of a pie-slice subregion of the entire space, which leads to the general expression for the corner contribution from a corner with angle $\alpha$. 

We start from (\ref{5-4b}) where the correlator is found in (\ref{4-21}). The indicator function for a pie-slice subregion with angle $\alpha$ is 
\begin{equation}
{\bf{1}}_{\vec{x} \in A} = e^{-\epsilon r^2} \Theta \left( \left( \frac{\alpha_i}{2} \right)^2 - \phi^2 \right)
\label{ind_crn_r}
\end{equation}
in polar coordinates. Here we have introduced a regulator $e^{-\epsilon r^2}$ with $\epsilon = 0+$ to avoid divergences at infinite radii. Thus, (\ref{5-4b}) in this particular case turns into
\begin{equation}
\begin{split}
\operatorname{tr} E_A
=&\frac{1}{\pi^2} \int_{-\alpha/2}^{\alpha/2} d \phi_1 \int_{\alpha/2}^{2\pi- \alpha/2} d\phi_2 \int_0^{\infty} d r_1 d r_2 ~r_1 r_2   \exp \left[ - (1+\epsilon) \left(r_1^2 + r_2^2\right) + 2r_1r_2 \cos (\phi_1-\phi_2) \right] \\
=&\frac{1}{\pi^2} \int_{-\alpha/2}^{\alpha/2} d \phi_1 \int_{-\pi+ \alpha/2}^{\pi- \alpha/2} d\phi_2 \int_0^{\infty} d r_1 d r_2 ~r_1 r_2   \exp \left[ - (1+\epsilon) \left(r_1^2 + r_2^2\right)  - 2r_1r_2 \cos (\phi_1-\phi_2) \right]
\end{split}
\label{v2c}
\end{equation}
We see that $\operatorname{tr} E_A$ is a function of $\alpha$ only. We obtain (\ref{6-1}) after multiplying by the factor $\frac{\pi^2}{3}$.

By utilizing the Jacobi-Anger identity we can integrate out $\phi_1$ and $\phi_2$ in (\ref{v2c}) first
\begin{equation}
\begin{split}
\operatorname{tr} E_A
=& -\frac{8}{\pi^2} \sum_{n = 1}^{\infty} (-i)^n \frac{ \sin^2 \frac{\alpha n}{2}}{n^2}\int_0^{\infty} d r_1 ~r_1 e^{-(1+\epsilon)r_1^2}  \int_0^{\infty} d r_2  ~r_2 e^{-(1+\epsilon)r_2^2}  J_n \left( i 2r_1 r_2 \right) \\
&+ \frac{1}{\pi^2} \alpha (2\pi -\alpha) \int_0^{\infty} d r_1  ~r_1 e^{-(1+\epsilon)r_1^2}  \int_0^{\infty} d r_2  ~r_2 e^{-(1+\epsilon)r_2^2}  J_0 \left( i 2r_1 r_2 \right)  \\
\end{split}
\end{equation}
and then the radii $r_1$ and $r_2$
\begin{equation}
\begin{split}
\operatorname{tr} E_A
=& -\frac{2}{\pi^2} \sum_{n = 1}^{\infty}  \frac{ \sin^2 \frac{\alpha n}{2}}{n^2} \Gamma \left( \frac{n}{2} +1 \right) \Gamma \left( \frac{n}{2} +1 \right) (1+\epsilon)^{-({2+n})} {{}_{2}{\mathbf{F}}_{1}}\left(\frac{n}{2}+1,\frac{n}{2}+1;n+1;\frac{1}{(1+\epsilon)^2} \right)  \\
&+ \frac{1}{4 \pi^2} \alpha (2\pi -\alpha) (1+\epsilon)^{-1} \int_0^{\infty} d \rho e^{-(1+\epsilon)\rho} e^{\frac{\rho}{1+\epsilon}} \\
\end{split}
\end{equation}

Given that $\epsilon = 0+$ we can expand the hypergeometric function around $1$ and keep the terms upto order $1$
\bea\label{angle_1}
\operatorname{tr} E_A
&= & -\frac{2}{\pi^2} \sum_{n = 1}^{\infty}  \frac{ \sin^2 \frac{\alpha n}{2}}{n^2}  (1+\epsilon)^{-2} \left( 1-\frac{1}{(1+\epsilon)^2}  \right)^{-1} (1+\epsilon)^{-n} \cr
&&+ \frac{1}{4 \pi^2} \alpha (2\pi -\alpha) (1+\epsilon)^{-2} \left( 1-\frac{1}{(1+\epsilon)^2}  \right)^{-1} \cr
&& -\frac{2}{\pi^2} \sum_{n = 1}^{\infty}  \frac{ \sin^2 \frac{\alpha n}{2}}{n^2} \left(\frac{n}{2}\right)^2 (1+\epsilon)^{-(2+n)} \left[ 2\psi \left( \frac{n}{2} +1 \right) + \log \left( 1-\frac{1}{(1+\epsilon)^2}  \right) + 2\gamma_E -1 \right] \cr
&& + \mathcal{O}(\epsilon \log \epsilon) \cr
&=& -\epsilon^{-1} \frac{1}{4\pi^2} \left[ 4 \sum_{n = 1}^{\infty}  \frac{ \sin^2 \frac{\alpha n}{2}}{n^2} +\frac{1}{2} \alpha (2\pi -\alpha)  \right] + \frac{1}{4\pi^2} \sum_{n = 1}^{\infty}  \frac{ 4\sin^2 \frac{\alpha n}{2}}{n}   \cr
&& - \left[  \log 2\epsilon  -1 \right]\frac{1}{8\pi^2} \sum_{n = 1}^{\infty}  4\sin^2 \frac{\alpha n}{2}    \cr
&& -\frac{1}{4\pi^2} \sum_{n = 1}^{\infty}  4 \sin^2 \frac{\alpha n}{2} \times  \left[ \psi \left( \frac{n}{2} +1 \right) +\gamma_E \right]+ \mathcal{O}(\epsilon \log \epsilon)
\eea
where $\psi (x)$ is the digamma function and $\gamma_E$ is the Euler-Mascheroni constant. Their sum can be expressed in terms of the integral
\begin{equation}
\psi\left(z\right)+\gamma_E=\int_{0}^{1}\frac{1-t^{z-1}}{1-t}\mathrm{d}t
\end{equation}
To simplify (\ref{angle_1}) we introduce polylogarithms to represent the series. Then we obtain
\begin{equation}
\begin{split}
\operatorname{tr} E_A
=& -\epsilon^{-1} \frac{1}{4\pi^2} \left[ \left( - \operatorname{Li}_2 (e^{i\alpha}) - \operatorname{Li}_2 (e^{-i\alpha}) + 2 \operatorname{Li}_2 (1) \right) +\frac{1}{2} \alpha (2\pi -\alpha)  \right] \\
&+ \frac{1}{4\pi^2} \left( - \operatorname{Li}_1 (e^{i\alpha}) - \operatorname{Li}_1 (e^{-i\alpha}) + 2 \operatorname{Li}_1 (1) \right)    \\
& - \left[  \log 2\epsilon  -1 +2 \int_{0}^{1}\frac{1}{1-t}\mathrm{d}t \right]\frac{1}{8\pi^2}\left( - \operatorname{Li}_0 (e^{i\alpha}) - \operatorname{Li}_0 (e^{-i\alpha}) + 2 \operatorname{Li}_0 (1) \right)    \\
& + \frac{1}{4\pi^2} \int_{0}^{1}\frac{\mathrm{d}t}{1-t} \left( - \operatorname{Li}_0 (e^{i\alpha}\sqrt{t}) - \operatorname{Li}_0 (e^{-i\alpha}\sqrt{t}) + 2 \operatorname{Li}_0 (\sqrt{t}) \right) + \mathcal{O}(\epsilon \log \epsilon)  \\
\end{split}
\label{angle_2}
\end{equation}
Since $\alpha$ is real and more concretely $\frac{\alpha}{2\pi} \in [0,1)$, we can utilize
\begin{equation}
\operatorname {Li} _{n}(e^{2\pi ix})+(-1)^{n}\operatorname {Li} _{n}(e^{-2\pi ix})=-{(2\pi i)^{n} \over n!}B_{n}(x),
\end{equation}
where $B_n (x)$ is the Bernoulli polynomial,
and find
\begin{eqnarray}
 &&\operatorname{Li}_2 (e^{i\alpha}) + \operatorname{Li}_2 (e^{i\alpha}) = \frac{1}{2}\left[ -\alpha (2\pi-\alpha) +\frac{2\pi^2}{3} \right] \\
 && \operatorname{Li}_0 (e^{i\alpha}) + \operatorname{Li}_0 (e^{i\alpha})  =-1 
\end{eqnarray}
Moreover, given that $\operatorname{Li}_2 (1) = \frac{\pi^2}{6}$, we see that the first term in (\ref{angle_2}) vanishes. 

As for the other terms, we notice that
\begin{eqnarray}
&& \operatorname {Li} _{1}(z)=-\ln(1-z) \label{li1d} \\
 && \operatorname {Li} _{0}(z)={z \over 1-z}
\end{eqnarray}
and integrate over $t$ first. We find
\begin{equation}
\begin{split}
&  \int_0^1 \frac{dt}{1-t} \left( - \operatorname{Li}_0 (\sqrt{t} e^{i\alpha}) - \operatorname{Li}_0 (\sqrt{t} e^{-i\alpha}) + 2 \operatorname{Li}_0 (\sqrt{t}) \right) \\
=&  -2  \frac{\cos \alpha}{|\sin \alpha|} \tan^{-1} \left|\cot \frac{\alpha}{2} \right| -2 \log \left( 2\left| \sin \frac{\alpha}{2} \right| \right) + 2 \operatorname{Li}_0(1) -2  \operatorname{Li}_1(1)
\end{split}
\end{equation}
where we have rewritten $\frac{u}{u-1}\big|_{u \to 1}$ and $\log (1-u) \big|_{u \to 1}$ into polylogarithms $\operatorname{Li}_0(1)$ and $\operatorname{Li}_0(1)$, respectively. 
Thus, we obtain from (\ref{angle_2})
\begin{equation}
\begin{split}
\operatorname{tr} E_A
=
& \frac{1}{4\pi^2} \left( - \operatorname{Li}_1 (e^{i\alpha}) - \operatorname{Li}_1 (e^{i\alpha}) + 2 \operatorname{Li}_1 (1) \right)    \\
& - \frac{1}{8\pi^2} \left[  \log 2\epsilon  -1 -2 \operatorname{Li}_1 (1) \right]\left( 1 + 2 \operatorname{Li}_0 (1) \right)    \\
& + \frac{1}{4\pi^2}\left[  -2  \frac{\cos \alpha}{|\sin \alpha|} \tan^{-1} \left|\cot \frac{\alpha}{2} \right| -2 \log \left( 2\left| \sin \frac{\alpha}{2} \right| \right) + 2 \operatorname{Li}_0(1) -2  \operatorname{Li}_1(1) \right] \\
=
& - \frac{1}{8\pi^2} \left[  \log 2\epsilon  -3 -2 \operatorname{Li}_1 (1) \right]\left( 1 + 2 \operatorname{Li}_0 (1) \right)    \\
& - \frac{1}{4\pi^2}\left[ 1 + 2  \frac{\cos \alpha}{|\sin \alpha|} \tan^{-1} \left|\cot \frac{\alpha}{2} \right|   \right] \\
\end{split}
\label{angle_3}
\end{equation}
where in the last equation we have plugged in (\ref{li1d}).

In (\ref{angle_3}) we have separated $\operatorname{tr} E_A$ into two parts -- an $\alpha$-independent part and an $\alpha$-dependent part. When $\alpha= \pi -\delta$ where $\delta \to 0$, we have
\begin{equation}
\begin{split}
& - \frac{1}{4\pi^2}\left[ 1 + 2  \frac{\cos \alpha}{|\sin \alpha|} \tan^{-1} \left|\cot \frac{\alpha}{2} \right|   \right] 
= -\frac{1}{4\pi^2} \left[1- 2 \cot \delta \tan^{-1} \tan \frac{\delta}{2} \right]  \\
= & -\frac{1}{4\pi^2} \left[1- 2 \left( \frac{1}{\delta} -\frac{1}{3}\delta - \mathcal{O}(\delta^3) \right)  \frac{\delta}{2} \right] 
=-\frac{1}{4\pi^2} \left[\frac{1}{3}(\pi -\alpha)^2 +\mathcal{O}((\pi -\alpha)^4)  \right]
\end{split}
\end{equation}
It implies that when $\alpha = \pi$, only the first line in the RHS of the last equation of (\ref{angle_3}) remains, i.e.
\begin{equation}
\operatorname{tr} E_A
=- \frac{1}{8\pi^2} \left[  \log 2\epsilon  -3 -2 \operatorname{Li}_1 (1) \right]\left( 1 + 2 \operatorname{Li}_0 (1) \right), \qquad \epsilon = 0+
\label{c16}
\end{equation}
On the other hand, when $\alpha = \pi$, we obtain a halfspace subregion, the entanglement entropy of which is given in (\ref{55-4}) via a calculation in Cartesian coordinates in section \ref{halfspace}. We learn from (\ref{55-4}) that there is no corner contribution. In other words, the RHS of (\ref{c16}) corresponds to the perimeter of the halfspace. Furthermore, if we compare the pie-slice subregions with various $\alpha$'s, we find that the perimeter terms corresponding to them are identical. Thus, we can subtract the perimeter term, in particular (\ref{c16}), from $\operatorname{tr} E_A$ corresponding to a corner with angle $\alpha$ to obtain the corner contribution, i.e.
\begin{equation}
a (\alpha) =\frac{\pi^2}{3} \left[ \operatorname{tr} E_A \big|_{\alpha} - \operatorname{tr} E_A\big|_{\alpha = \pi} \right]= - \frac{1}{12}\left[ 1 + 2  \frac{\cos \alpha}{|\sin \alpha|} \tan^{-1} \left|\cot \frac{\alpha}{2} \right|   \right]
\tag{\ref{8-6}}
\end{equation}

For subregions with more than one corner, we can use (\ref{7-12}) to determine the contribution to entanglement entropy from the corners.

\end{document}